\documentclass[fleqn, usenatbib]{mnras}

\usepackage{newtxtext, newtxmath}

\usepackage[T1]{fontenc}

\DeclareRobustCommand{\VAN}[3]{#2}
\let\VANthebibliography\thebibliography
\def\thebibliography{\DeclareRobustCommand{\VAN}[3]{##3}\VANthebibliography}

\usepackage{graphicx}	
\usepackage{amsmath}	
\usepackage{hyperref}
\usepackage{orcidlink}

\usepackage{CJKutf8}
\usepackage{color, lipsum}
\usepackage[normalem]{ulem}  

\newcommand{\iso}[2]{{}^{#2}{\rm #1}}
\newcommand{\kanji}[1]{\begin{CJK}{UTF8}{ipxm}(#1)\end{CJK}}

\title[Uncertainties of iron-group nucleus in CCSNe]{Uncertainties in the production of iron-group nuclides in core-collapse supernovae from Monte Carlo variations of reaction rates}

\author[N.~Nishimura et al.]{
Nobuya~Nishimura \kanji{西村信哉}$^{1,2,3,4}\orcidlink{0000-0002-0842-7856}$\thanks{E-mail: nobuya@sin.cc.kogakuin.ac.jp, nobuya@cns.s.u-tokyo.ac.jp (NN)}, 
Carla Fr\"ohlich$^5$\orcidlink{0000-0003-0191-2477} 
and Thomas Rauscher$^{6,7}$\orcidlink{0000-0002-1266-0642}
\\
$^{1}$Academic Support Center (ASC), Kogakuin University, Hachioji, Tokyo 192-0015, Japan\\
$^{2}$Center for Nuclear Study (CNS), The University of Tokyo, Bunkyo-ku, Tokyo 113-0033, Japan\\
$^{3}$Astrophysical Big-Bang Laboratory, RIKEN Pioneering Research Institute (PRI), Wako, Saitama 351-0198, Japan\\
$^{4}$National Astronomical Observatory of Japan (NAOJ), Mitaka, Tokyo 181-8588, Japan\\
$^{5}$Department of Physics, North Carolina State University, Raleigh NC 27695, USA\\
$^{6}$Department of Physics, University of Basel, CH-4056 Basel, Switzerland\\
$^{7}$Centre for Astrophysics Research, University of Hertfordshire, Hatfield AL10 9AB, UK
}

\date{Accepted XXX. Received YYY; in original form ZZZ}

\pubyear{\the\year{}}

\begin{document}
\label{firstpage}
\pagerange{\pageref{firstpage}--\pageref{lastpage}}
\maketitle

\begin{abstract}
Core-collapse supernovae, occurring at the end of massive star evolution, produce heavy elements, including those in the iron peak. Although the explosion mechanism is not yet fully understood, theoretical models can reproduce optical observations and observed elemental abundances. However, many nuclear reaction rates involved in explosive nucleosynthesis have large uncertainties, impacting the reliability of abundance predictions. To address this, we have previously developed a Monte Carlo-based nucleosynthesis code that accounts for reaction rate uncertainties and has been applied to nucleosynthesis processes beyond iron. Our framework is also well suited for studying explosive nucleosynthesis in supernovae. In this paper, we investigate 1D explosion models using the ``PUSH method'', focusing on progenitors with varying metallicities and initial masses around $M_{\rm ZAMS} = 16 M_{\odot}$. Detailed post-process nucleosynthesis calculations and Monte Carlo analyses are used to explore the effects of reaction rate uncertainties and to identify key reaction rates in explosive nucleosynthesis. We find that many reactions have little impact on the production of iron-group nuclei, as these elements are primarily synthesized in the nuclear statistical equilibrium. However, we identify a few ``key reactions'' that significantly influence the production of radioactive nuclei, which may affect astrophysical observables. In particular, for the production of $\iso{Ti}{44}$, we confirm that several traditionally studied nuclear reactions have a strong impact. However, determining a single reaction rate is insufficient to draw a definitive conclusion.
\end{abstract}

\begin{keywords}
nuclear reactions, nucleosynthesis, abundance -- stars: neutron -- stars: abundances -- supernovae: general
\end{keywords}



\section{Introduction}

Massive stars exceeding a zero-age-main-sequence mass ($M_\mathrm{ZAMS}$) of $8$--$10M_{\odot}$ explode as a core-collapse supernova (CCSN) at the end of their evolution \citep[e.g.,][]{2003ApJ...591..288H, 2025MNRAS.543.2796H}. These explosions eject various elements ranging from helium to iron, produced during the long-term stellar evolution, and the iron-peak and lighter trans-iron nuclides produced by explosive nucleosynthesis of CCSNe \citep[e.g.,][]{1995ApJS..101..181W, 1996ApJ...460..408T, 2002ApJ...576..323R}. Nucleosynthesis products from CCSNe are not only a primary source of metal enrichment in the Galactic chemical evolution \citep[e.g.,][]{2006ApJ...653.1145K, 2020ApJ...900..179K} but also the cosmic source of several radioactive nuclei. These radioactive nuclei, predominantly those up to iron, exhibit a wide range of decay half-lives and can be directly identified through astrophysical observations. In the central engine of the SN explosion a substantial amount of iron-group radioactive nuclei is synthesized, with $\iso{Ni}{56}$ being the most prominent and primary observable as a heating source for the early phase of CCSNe light curve. The decay $\iso{Ni}{56} \rightarrow \iso{Co}{56}$ (the half~life $\tau_{1/2} \sim 6.081~{\rm d}$) and its subsequent decay (i.e., $\iso{Co}{56} \rightarrow \iso{Fe}{56}$ with $\tau_{1/2} \sim 77.24~{\rm d}$) is closely linked to temporal evolution of the optical light curves. Additionally, longer-lived nuclei such as $\iso{Ti}{44}$ ($\tau_{1/2} \sim 59.1~{\rm y}$) are detectable in SN remnants even centuries after the explosion \citep{1994A&A...284L...1I, 2014Natur.506..339G, 2015Sci...348..670B, 2015A&A...579A.124S, 2017ApJ...842...13W}, as observed in several historical SNe. Moreover, $\gamma$-ray emissions from nuclides like $\iso{Al}{26}$, with a half-life of $\tau_{1/2} \sim 0.717~{\rm Myr}$), enable the observation of nucleosynthesis products across the Galaxy \citep[e.g.,][]{2006Natur.439...45D}\footnote{Half-life values in this paragraph are adopted from ``Nuclear Wallet Cards'' 2023 edition (https://www.nndc.bnl.gov).}.

Although the explosion mechanism of CCSNe driven by neutrino heating -- in which complex 3D convective motion around the proto-neutron star (NS) is involved \citep[for reviews, e.g.,][]{2013RvMP...85..245B, 2016ARNPS..66..341J, 2024PJAB..100..190Y, 2025arXiv250214836J} -- is not fully understood, spherically-symmetric explosion models have been constructed that relatively well reproduce the observed elemental abundances. Physically simplified, artificial 1D explosion models are also used to justify 3D models and to examine observational constraints \citep[see e.g.,][for recent discussion on the validity of 1D modeling]{2012ApJ...757...69U, 2015ApJ...806..275P, 2014ApJ...782...91N, 2016ApJ...821...69E, 2019ApJ...870....1E, 2022ApJ...929...43G, 2023arXiv230103610S, 2023MNRAS.518.1818I, 2025arXiv250113172I}. Such models are ideal to systematically study the impact of microphysics, e.g., the equation of state and weak interactions, on CCSNe and nuclear reaction rates on nucleosynthesis. Some of the key nuclear reactions in explosive nucleosynthesis, most of which involve radioactive nuclei, can be accessible through accelerator experiments, offering the potential to experimentally investigate unmeasured reaction rates and relevant nuclear properties with importance to astrophysics.

Explosive nucleosynthesis in CCSNe is significantly influenced by uncertainties in explosion models and nuclear-physics inputs. Even small variations within a single model can substantially impact galactic chemical evolution scenario, while differences in estimates of key radioactive elements can affect interpretation of direct optical observations and impose constraints on explosion models. Systematic studies of explosion models and nuclear reaction rates have been conducted for the production of nuclei such as $\iso{Ni}{56}$ and $\iso{Ti}{44}$. For explosion models, there is a range of cases, from simplified one-dimensional models to multi-dimensional models that aim to be as self-consistent as possible, partly due to the incomplete understanding of the explosion mechanism. Regarding the elements produced, the temperature and density (entropy) environment, along with the dependence on physical parameters such as the electron abundance $Y_e$, are generally well understood, except for the complexity introduced by explosion dynamics. In addition to the complexity of the explosion model, uncertainties from multiple nuclear reactions and decays propagate nonlinearly along nucleosynthesis flows. This makes it difficult to identify key uncertain reactions using traditional systematic impact studies that vary a single reaction rate at a time.

To quantify the impact of the uncertainty in the nuclear reaction rates, it is useful to study it numerically and to compare the results with observed quantities through nuclear reaction network calculations. This is traditionally done by manual variation of rates, using a fixed variation factor that accounting for the rate uncertainty for individual rates or groups of rates \citep[see, e.g.,][]{1998ApJ...504..500T, 1999ApJ...521..735H, 2002ApJS..142..105I, 2007ApJ...671..821T, 2010ApJ...715.1383H, 2020ApJ...901...77H, 2020ApJ...898....5S}. In recent years, there has been an increasing number of studies using Monte Carlo (MC) variations of reaction rates in a variety of stellar environments \citep[see, e.g.,][]{2018ApJS..234...19F, 2021MNRAS.503.3913D, 2024A&A...684A...8M}. There are several advantages of the MC approach. A \textit{simultaneous} variation of many rates with many different, individual variation factors samples the overall uncertainty in a produced abundance stemming from all contributing rates and not only for the selected group of rates. Moreover, the importance of a specific rate in the ensemble of contributing rates is better assessed. A simple, individual variation of single rates (or group of rates using the same variation factor) shows the dependence on an abundance to this rate but may overestimate its contribution to the total production uncertainty of a specific nuclide.

We have developed a nucleosynthesis code \citep{2016MNRAS.463.4153R, 2017MNRAS.469.1752N} within a Monte Carlo (MC) framework including a nuclear reaction network solver and statistical analysis tools. We have already applied it to several processes beyond iron \citep{2016MNRAS.463.4153R, 2017MNRAS.469.1752N, 2018MNRAS.474.3133N, 2018MNRAS.478.4101C, 2019MNRAS.489.1379N}. Given its general applicability, our framework is naturally suited for studying explosive nucleosynthesis in SNe. Although here we focus on the production of short-lived nuclei not considered in our previous studies, the application of our MC nucleosynthesis framework to this case is technically straightforward.

In this study, we investigate 1D explosion models using explosive tracers obtained with the {\it PUSH method}, which triggers explosions by mimicking the enhanced neutrino heating observed in multi-dimensional simulations. We focus on SN explosive nucleosynthesis in progenitors with solar and sub-solar metallicities and an initial mass of $M_{\rm ZAMS} = 16,M_{\odot}$. Detailed post-process nucleosynthesis calculations with MC analysis are employed to comprehensively explore the effects of uncertainties in relevant reaction rates. Additionally, we discuss key reaction rates for explosive nucleosynthesis based on a statistical analysis of our MC nucleosynthesis results.

The paper is organized as follows: Methods are described in Section~\ref{sec:methods}, and the overall results for key reactions presented in Section~\ref{sec:results}. We especially focus on selected nuclides important for observations. Section~\ref{sec:conclusion} is dedicated to the summary and conclusion.


\section{Methods}
\label{sec:methods}

\subsection{Monte Carlo Nucleosynthesis}
\label{sec:mcmethod}
\begin{figure}
\includegraphics[width=\columnwidth]{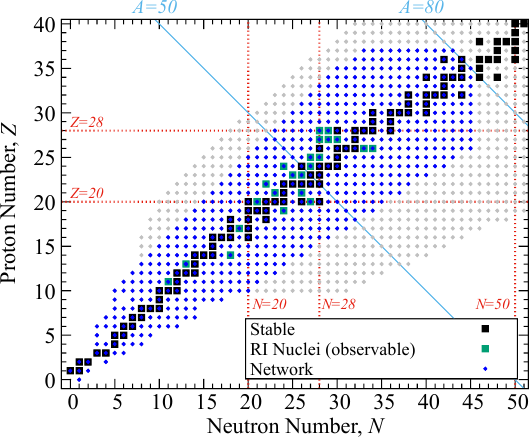}
\caption{The $615$ nuclei contained in the network (blue circles), overlaid on all nuclei within the drip lines (grey circles) as determined by the FRDM mass model \citep[][]{1995ADNDT..59..185M}. Among them, the $90$ stable nuclei and $24$ selected radioactive species (see Table~\ref{tab:ri_nuc}) are indicated by black and green squares, respectively.}
\label{fig:network}
\end{figure}

\begin{table}
\caption{Key radioactive nuclei considered in this study.}
\label{tab:ri_nuc}
\begin{tabular}{ccccccccccc}
\hline
Nuclide ($Z$) & $A$ & Nuclide & $A$ & Nuclide & $A$\\
\hline
Na ($11$) & 22 & Ca ($20$) & $41, 47$ & Mn ($25$) & 52, 53 \\
Al ($13$) & 26 & Sc ($21$) & $44, 47$ & Fe ($26$) & 59, 60 \\
Si ($14$) & 32 & Ti ($22$) & $44$     & Co ($27$) & 55, 56, 57 \\
Cl ($17$) & 36 & V  ($23$) & $48, 49$ & Ni ($28$) & 56, 57, 59 \\
K  ($19$) & 43 & Cr ($24$) & $48, 51$ & & \\
\hline
\end{tabular}
\end{table}

\begin{table}
\caption{The upper and lower limits of theory uncertainty factors used for various reaction types.}
\label{tab:unc}
\begin{tabular}{lccccccccccc}
\hline
Reaction & $({\rm n},\gamma)$ & $({\rm p},\gamma)$ & $({\rm p},{\rm n})$ & $(\alpha,\gamma)$ & $(\alpha,{\rm n})$ & $(\alpha,{\rm p})$\\
\hline
$U^{\rm up}$  & 2 & 2 & 2 & 2 & 2 & 2\\
$U^{\rm low}$ & $1/2$ & $1/3$ & $1/3$ & $1/10$ & $1/10$ & $1/10$\\
\hline
\end{tabular}
\end{table}

\begin{figure*}
\centering

\includegraphics[width=\hsize]{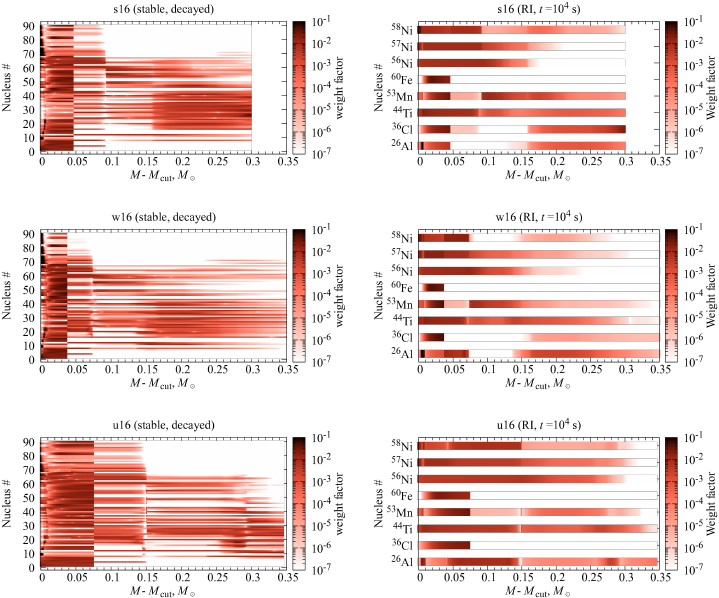}

\caption{Weight factors of nuclei after decay for the models defined in Section~\ref{sec:models} and summarised in Table \ref{tab:models}, shown for s16 (top), w16 (middle), and u16 (bottom). The left panels show stable nuclei and key radioactive nuclei with $A \lesssim 80$, while the right panels display results for 8 selected key radioactive nuclei. The final abundances of stable nuclei are taken from the values after all decays, while those of radioactive nuclei are taken at the end of the nucleosynthesis calculations ($t = 10^4~{\rm s}$). The order of adopted stable nuclides is p, d, $\iso{He}{3,4}$, $\iso{Li}{7}$, $\iso{B}{10,11}$, $\iso{C}{12,13}$, $\iso{N}{14,15}$, $\iso{O}{16,17,18}$, $\iso{F}{19}$, $\iso{Ne}{20,21,22}$, $\iso{Na}{23}$, $\iso{Mg}{24,25,26}$, $\iso{Al}{27}$, $\iso{Si}{28,29,30}$, $\iso{P}{31}$, $\iso{S}{32,33,34,36}$, $\iso{Cl}{35,37}$, $\iso{Ar}{36,38,40}$, $\iso{K}{39,40,41}$, $\iso{Ca}{40,42,43,44,46,48}$, $\iso{Sc}{45}$, $\iso{Ti}{46,47,48,49,50}$, $\iso{V}{50,51}$, $\iso{Cr}{50,52,53,54}$, $\iso{Mn}{55}$, $\iso{Fe}{54,56,57,58}$, $\iso{Co}{59}$, $\iso{Ni}{58,60,61,62,64}$, $\iso{Cu}{63,65}$, $\iso{Zn}{64,66,67,68,70}$, $\iso{Ga}{69,71}$, $\iso{Ge}{70,72,73,74,76}$, $\iso{As}{75}$, and $\iso{Br}{79}$.}
\label{fig:weight}
\end{figure*}

\begin{figure}
\centering

\includegraphics[width=\columnwidth]{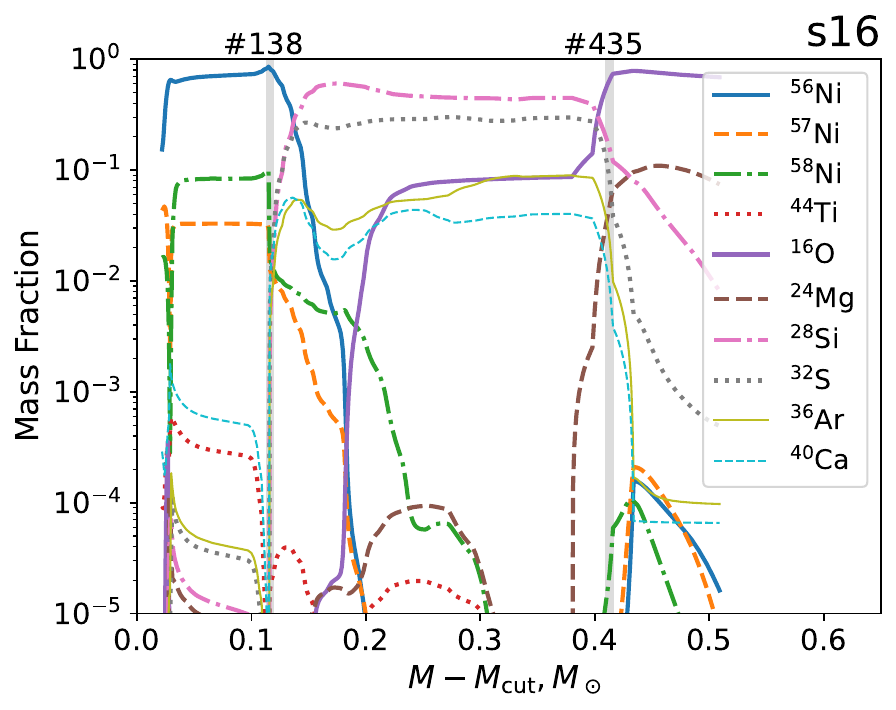}
\includegraphics[width=\columnwidth]{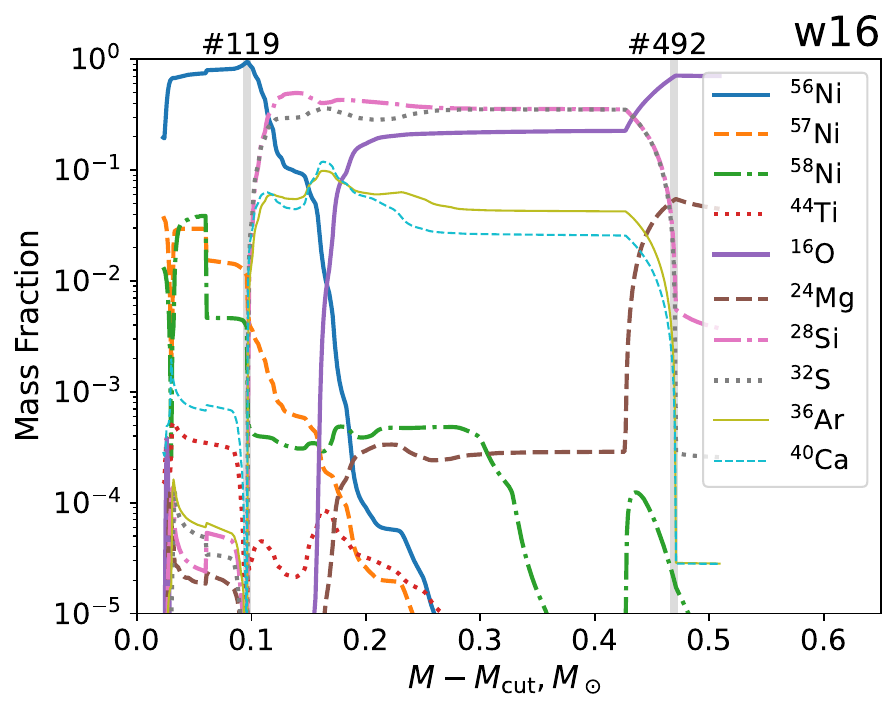}
\includegraphics[width=\columnwidth]{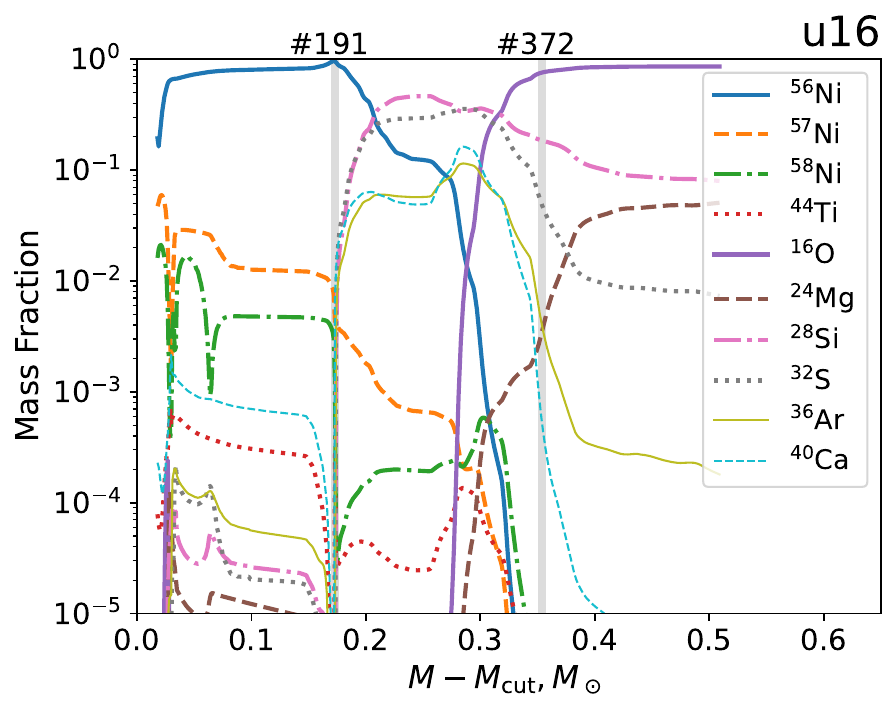}

\caption{Nucleosynthesis yields of explosive nucleosynthesis at $10^4$~s after explosion for $16 M_\odot$ stellar models. The s16 and w16 progenitors have solar abundance, whereas u16 is based on a metal poor star (see Table~\ref{tab:models}). Vertical gray bands labeled with zone numbers indicate the boundaries of the Si-, O-, and Ne-burning layers: $0.116$ and $0.413~M_\odot$ for s16,  $0.096$ and $0.469~M_\odot$ for w16, and  $0.173$ and $0.354~M_\odot$ for u16 in the horizontal axis ($M-M_{\mathrm{cut}}$).}
\label{fig:std_nucl}
\end{figure}

We post-process tracers -- each representing the time-evolution of temperature and density in a mass zone of the exploding star -- and follow explosive nucleosynthesis with the \textsc{PizBuin} MC framework \citep{2016MNRAS.463.4153R}, containing a nuclear reaction-network code \textsc{MC-WinNet}. This code is based on a previous version of \textsc{WinNet} \citep{2023ApJS..268...66R}, extended to incorporate rate variations and parallelized using OpenMP\footnote{The older version of \textsc{WinNet} is not thread-safe, so caution is required when running \textsc{WinNet} in parallel systems.} \citep[see, e.g.,][]{2016MNRAS.463.4153R}. Basic properties, methods for rate variation and analysis techniques are described in \cite{2016MNRAS.463.4153R} with application to SN explosive nucleosynthesis, including the p-process. Treatment of $\beta$-decay uncertainties and applications to the s-process are provided in \cite{2017MNRAS.469.1752N}. The methodology for multi-zone SN explosion models is shown in \cite{2016MNRAS.463.4153R} and \cite{2018MNRAS.474.3133N}.

In the present study, we adopt a nuclear reaction network that includes $615$ nuclides with a total of $7,992$ nuclear decays and reactions (distinguished forward and reverse reactions), ranging from nucleons to heavier nuclides with a mass number of up to $A = 80$. This selection effectively covers iron-group nuclides, as shown in Figure~\ref{fig:network}, while minimizing the number of nuclear species considered to enhance computational efficiency. The network is not extended to the theoretical drip lines for mid-heavy nuclei but is instead restricted to nuclides that can be significantly produced by explosive nucleosynthesis in our calculations. An upper mass limit of $A = 80$ is sufficient for our current purposes, since all key nuclei and reactions lie below $A \sim 60$. In the present study, we analyse nuclear reaction and decay uncertainties for the production of 90 nucleus stable and 24 selected radioactive species, listed in Table~\ref{tab:ri_nuc} for latter. We adopt the standard reaction and decay rates from the latest JINA Reaclib Database (``default'', revised from Ver.~2.2 on 2021-06-24) \citep{2000ADNDT..75....1R, 2010ApJS..189..240C},\footnote{JINA Reaclib: https://reaclib.jinaweb.org} which compiles experimentally available reaction rates along with theoretical predictions, the latter being particularly necessary for unstable nuclides.

We apply temperature-dependent uncertainties to each reaction rate, with the uncertainty factors defined as functions of temperature, arising from a combination of experimentally known ground-state rates and predictions from rates on thermally excited target nuclides \citep[see,][]{2016MNRAS.463.4153R, 2017MNRAS.469.1752N}. However, this approach is applicable only to reaction rates with experimentally determined ground-state contributions \citep{2011ApJ...738..143R, 2012ApJS..201...26R, 2020nnc..confa0061R}. In explosive nucleosynthesis, most reaction rates in the network are theoretically derived. For purely theoretical rates, we adopt a constant uncertainty factor across all temperature values. Starting from the standard reaction rate values, we introduce variations by defining lower and upper limits, as listed in Table~\ref{tab:unc}. The choice of these limits is discussed in detail in Sec.~2.3.2 of \citet{2016MNRAS.463.4153R}. For theoretically predicted rates, the uncertainty range implemented by these limits is considered to comprise uncertainties arising from predicted nuclear properties and from the applicability of the reaction models themselves. Thus, this is a model-independent approach. Within these limits, we vary the reaction rates using a uniform random distribution within the specified uncertainty range.

We note that the forward and reverse reaction pairs (e.g., A$({\rm n},\gamma)$B and B$(\gamma,{\rm n})$A) should not be varied independently. For these reaction pairs, the reaction rates must be adjusted in a physically consistent manner to satisfy the principle of detailed balance.\footnote{The \textsc{REACLIB} format \citep{2000ADNDT..75....1R}, with its seven fitted parameters, has the advantage to implement and ensure detailed balance, as the reverse reaction is formulated based on the forward reaction.} We therefore denote the reaction pair as, e.g., ${\rm A} + {\rm n} \leftrightarrow \gamma + {\rm B}$ \citep[see also, e.g.,][]{2016MNRAS.463.4153R, 2017MNRAS.469.1752N}. Consequently, the uncertainties in the reaction rates do not affect the abundances when the network is in equilibrium (nuclear statistical equilibrium, NSE) \citep{2020entn.book.....R}. The NSE abundances are sensitive to the nuclear binding energies but nuclear masses are well known for the considered range of nuclides and thus \textit{nuclear} uncertainties provide a negligible contribution to the overall uncertainties of NSE abundances. We adopt a 30\% uncertainty for weak-decay rates, i.e., $\beta$ decay and $\mathrm{e}^{\pm}$-capture half-lives, corresponding to a factor of $1.3$ for the upper limit and $1/1.3$ for the lower limit. This choice is motivated by the fact that all relevant half-lives are based on experimental evaluations in the context of explosive nucleosynthesis. Following the notation of Nuclear Wallet Cards database\citep[adopted by JINA Reaclib:][]{2010ApJS..189..240C}, we denote $\varepsilon+\beta^+$ as the sum of $\mathrm{e}^-$ capture and $\beta^+$ decay, while $\varepsilon$ refers to the case of $\mathrm{e}^-$ capture only. We note that, unlike in our s-process study \citep{2017MNRAS.469.1752N}, we do not include temperature-dependent uncertainties for $\beta$ decay.

Although the nucleosynthesis details for the PUSH multi-zone models differ from those in our previous MC studies, the approach itself is identical to that used in our previous work. In particular, it is comparable to applications of the $\gamma$-process in 1D multi-zone CCSN models \citep{2016MNRAS.463.4153R} and multi-dimensional (2D) Type Ia SN models \citep{2018MNRAS.474.3133N}. Our procedure is carried out in three steps: (i) Perform standard nucleosynthesis calculations without varying the reaction rates and determination of a weight factor for each zone based on the amount of each nuclide produced (or depleted); (ii) conduct independent MC nucleosynthesis calculations for all zones; (iii) Determine the uncertainty range in the weighted average of nucleosynthesis products using the results from all zones combined utilizing the weight factors and identify key reactions through statistical analysis.

\begin{table*}
\caption{Properties of the $16 M_\odot$ progenitor and explosion models.}
\label{tab:models}
\begin{tabular}{lccccccccccccccc}
\hline
Model & Metallicity & \# of Progenitor & \# of Mass- & O-burning & $E_{\mathrm{expl}}$ & $M_{\mathrm{cut}}$ & Layer of $M_{\mathrm{cut}}$${}^{c}$ & Explosive Si& $^{56}$Ni mass\\
      &             & Nuclides & coordinate for MC ${}^{a}$ & Core Mass${}^{b}$ & &  &  & burning Zones${}^{d}$ & (post process)\\
 & $Z_{\odot}$ & & $M_\odot$ (zone) & $M_\odot$ (zone \#) & $10^{51}$erg & $M_{\odot}$ & & & $M_{\odot}$ \\
\hline
s16 & $1$       &   20 & $1.845$ (\#22--321) & 1.959 (\#435) & 1.365 & 1.524 & Si-O & \#22--138 &$0.2621$ \\
w16 & $1$       & 1505 & $1.886$ (\#23--372) & 2.001 (\#492) &1.256 & 1.514 & Si-O & \#23--119 &$0.2047$ \\
u16 & $10^{-4}$ &   20 & $2.110$ (\#18--367) & 2.115 (\#372) &1.868 & 1.743 & Si-O & \#18--191 &$0.4307$ \\
\hline
\end{tabular}
\begin{flushleft}
$^{a}$ The enclosed mass and total number of zones included in the MC calculations, in which explosive nucleosynthesis occurs.\\
$^{b}$ The enclosed mass of the oxygen-burning core and the corresponding zone. Note that this exceeds the region considered in the MC calculations.\\
$^{c}$ Mass corresponding to a remnant NS, which is omitted in this study.\\
$^{d}$ Complete silicon burning is achieved, reaching NSE in which $\iso{Ni}{56}$ is the predominant nucleus.
\end{flushleft}
\end{table*}

The statistical analysis was implemented as an automated procedure to identify the most important reactions in complex flow patterns from superposition of many zones (tracers). Such an approach is more feasible and superior to visual inspection of flows and manual variation of limited rate sets, especially when a large number of reactions, nuclides, and abundances is involved. It is based on examining correlations between variations in rates and abundances. Various definitions of correlation identifiers are found in literature and can be categorized into rank methods and product-moment methods \citep{1955kendall,2017hemmerich,2020nnc..confa0061R}. Rank correlation methods, although they are formally assumed to better account for data outliers, are losing information in the ranking procedure and thus are rather unsuited for the purpose of correlating reactions and abundances \citep{1955kendall,2017hemmerich}. Therefore the more suitable Pearson product-moment correlation coefficient was chosen to quantify correlations \citep{1895RSPS...58..240P}. This is safe because data outliers, to which the Pearson coefficient would be vulnerable, do not appear in an analytical, limited variation of reaction rates. Moreover, the Pearson coefficient is simpler to handle, especially when calculating a combined, weighted correlation including many tracers (see below). Positive values of the Pearson coefficient $-1\leq r_\mathrm{corr} \leq 1$ indicate a direct correlation between rate change and abundance change, whereas negative values signify an inverse correlation, i.e., the abundance decreases when the rate is increased. Larger absolute values $|r_\mathrm{corr}|$ indicate a stronger correlation and this can be used for extracting the most important reactions from the MC data. Consistent with our previous studies, we define a key rate (i.e., a rate dominating the final uncertainty in the production of a nuclide) by having $|r_\mathrm{corr}|\geq 0.65$.

As detailed further below, we aimed to identify important reactions within single zones but also key reactions globally affecting the final abundances. For the latter, it was necessary to modify the basic Pearson formula to provide a weighted average over all zones. This avoids overemphasizing contributions from zones where the respective reactions do not significantly affect any abundances. Our weighted correlation coefficient $\overline{r}_\mathrm{corr}(q,z)$ for the abundance $^qY$ of nuclide $q$ with respect to reaction $z$ is given by \citep{2016MNRAS.463.4153R,2020JPhCS1643a2062R}
\begin{equation}
\label{eq:weightcorr}
\overline{r}_\mathrm{corr}(q,z)=\frac{\sum_{ij} {^qw_{j}^2} \left( {^zf_{ij}} -\overline{f}_j \right)\left( ^qY_{ij} - \overline{^qY}_j\right) 
}{\sqrt{\sum_{ij} {^qw_j^2}\left( {^zf_{ij}} -\overline{f}_j \right)^2}\sqrt{\sum_{ij} {^qw_j^2} \left( ^qY_{ij} -\overline{^qY}_j 
\right)^2}} \ .
\end{equation}
The zone is identified by the tracer number $j$ and the iteration by $i$, with a variation factor $^zf_{ij}= {^zr^*_{\mathrm{MC},ij}} / {^zr^*_{\mathrm{std},ij}}$ of the rate of reaction $z$ and the final abundance $^qY_{ij}$ of nuclide $q$ resulting from this variation. The barred quantities are the means of the samples of variation factors $\overline{f}_j=\left(\sum_{i=1}^k {^zf_{ij}} \right)/k$ and abundances $\overline{^qY}_j=\left(\sum_{i=1}^k {^qY_{ij}} \right)/k$ with respect to the number of MC iterations $k$. As in our previous work, $k=10000$. The standard library rate is denoted by $^zr^*_\mathrm{std}$ and $^zr^*_\mathrm{MC}$ is the rate after multiplying the standard rate by the MC variation factor. The asterisks imply that these are \textit{stellar} rates, i.e., including reactions on thermally excited states of the involved nuclides. This is especially important at the elevated temperatures encountered in explosive burning.

To connect all rates to all abundances of 
interest, the number of weighted correlation factors to be computed for each nuclide of interest is the number of reactions in the 
network. The weight of each zone is calculated from the relative abundance change
\begin{equation}
\label{eq:normprod}
^qw_j=\frac{|{^qY^\mathrm{std}_j}-{^qY^\mathrm{ini}_j}|}{\sum_j {|{^qY^\mathrm{std}_j}-{^qY^\mathrm{ini}_j}}|}
\end{equation}
for each nuclide $q$ with initial abundance $^qY^\mathrm{ini}_j$ in trajectory $j$. The final abundances obtained with the standard rate set are denoted by $^qY^\mathrm{std}_j$. The obtained weight factors are shown in Figure~\ref{fig:weight}. These also illustrate the regions of the star where the abundance of a nuclide is dominantly affected.

Note that the final abundances of stable nuclei are taken after all decays, whereas those of radioactive nuclei are evaluated at the end of the nucleosynthesis calculations ($t = 10^4~{\rm s}$). In the present study, weight factors are computed and applied separately for stable and unstable nuclei at their respective evaluation times. Consequently, decay processes occurring well after $10^4~\mathrm{s}$ are beyond the scope of the present scheme. Nevertheless, such late-time decays are mostly well known, for example, the decay chain $\iso{Ni}{56} \rightarrow \iso{Co}{56} \rightarrow \iso{Fe}{56}$, whose half lives much longer than one day. The results presented in the Section~\ref{sec:results} concern not only weighted correlations but also correlations for individual zones. Obviously, in the latter case $j=1$. Then the weights and variation factors reduce to $^qw_j={^qw_1}=1$ and ${^zf_{ij}}={^zf_{i1}}={^zf_{i}}={^zr^*_{\mathrm{MC},i}}/{^zr^*_{\mathrm{std},i}}$, respectively.

\subsection{SN Explosion models}
\label{sec:models}

In the present study, we adopt SN explosion models by the PUSH method. The detailed methods and results are shown in the series of papers; Paper~I--V \citep{2015ApJ...806..275P, 2019ApJ...870....1E, 2019ApJ...870....2C, 2020ApJ...888...91E, 2022ApJ...929...43G}. Here, we adopt simulation models from Paper~V \citep{2022ApJ...929...43G}, which discusses the dependence on the nuclear equation of state (EOS) and stellar metallicity. The relevant features of the PUSH method for explosive nucleosynthesis are that the electron fraction is consistently evolved throughout the infall, bounce, and expansion of the material. In addition, the separation between ejecta and the forming proto-neutron star emerges consistently from the simulation, i.e., we do not have to impose a mass cut by hand.

The specific explosion models adopted in the present work are s16, w16, and u16 exploded using the SFHo EOS \citep{2013ApJ...774...17S} based on $16 M_\odot$ progenitors \citep{2022ApJ...929...43G}, corresponding to s16.0, w16.0, and u16 with SFHo, respectively, in the original paper. Table~\ref{tab:models} summarizes the key characteristics of the three models, covering progenitor properties, explosion dynamics, and nucleosynthesis products. The column ``\# of Progenitor Nuclides'' refers to the number of nuclides in the progenitor model file, not to the size of the extended network for the MC post-processing (shown in Figure~\ref{fig:network}). The masses of $\iso{Ni}{56}$ are taken from the end of the present post-processing calculations at $t = 10^4~{\rm s}$.

These models represent stars of 16~M$_{\odot}$ zero-age main sequence mass at two different metallicities (solar metallicity and $10^{-4}$ of solar metallicity). This is primarily because differences in explosive nucleosynthesis have been shown to vary significantly with changes in metal abundances during stellar evolution. Both the s16 and w16 models assume solar metallicity at the birth of the star, with the primary difference arising from the numerical treatment of stellar evolution. The w16 model includes a more extensive reaction network, incorporating more nuclides. The u16 model represents a more metal-poor star with a lower initial metallicity. Such stars have significantly different stellar structures, resulting in notably distinct outcomes. Although nucleosynthesis can occur at the same peak temperature, the u16 model has almost twice as much ejecta mass reaching conditions typical for $^{56}$Ni production. In combination, these three models allow to investigate a range of explosion conditions. 

The final abundances at $10^4~{\rm s}$ after explosion from post-processing with our network code are shown in Figure~\ref{fig:std_nucl}. Unless otherwise specified, final abundances are defined as fully decayed values for stable nuclei and as the values at $t = 10^4~{\rm s}$ after the explosion for radioactive nuclei. These are typical explosive nucleosynthesis results and are consistent with the original results in \cite{2022ApJ...929...43G} within differences of input physics such as reaction rates in the nuclear reaction network. It is to be noted that we did not include neutrino-induced reactions in the post-processing network and therefore do not include the $\nu$p process \citep{2006ApJ...637..415F,2006ApJ...644.1028P,2006ApJ...647.1323W} products. Throughout the remainder of this paper, unless otherwise specified, we define the final abundances as the fully decayed values for stable nuclei and as the values at $t = 10^4~{\rm s}$ after the explosion for radioactive nuclei. Under this definition, the final abundances of stable nuclei differ slightly from those shown in Figure~\ref{fig:std_nucl}.


\section{Results}
\label{sec:results}

In Section~\ref{res:mc-full}, as in our previous work, we provide an uncertainty assessment and identify ``key reactions'' for the supernova explosion models as a whole, considering all relevant zones and the total nucleosynthesis by integrating over the different stellar layers (see Table~\ref{tab:models} for the included zones). In Section~\ref{res:mc-layer}, we present the same analysis focusing on the layer of explosive silicon burning, where the nucleosynthesis products are distributed relatively homogeneously within the zone (see Figure~\ref{fig:std_nucl}). In Section~\ref{sec:obs}, we discuss the abundance uncertainties and key reactions associated with radioactive nuclei of interest for optical observations.

\subsection{Abundance uncertainty and key reactions: general features}
\label{res:mc-full}

\begin{figure}
\centering
\includegraphics[width=1\hsize]{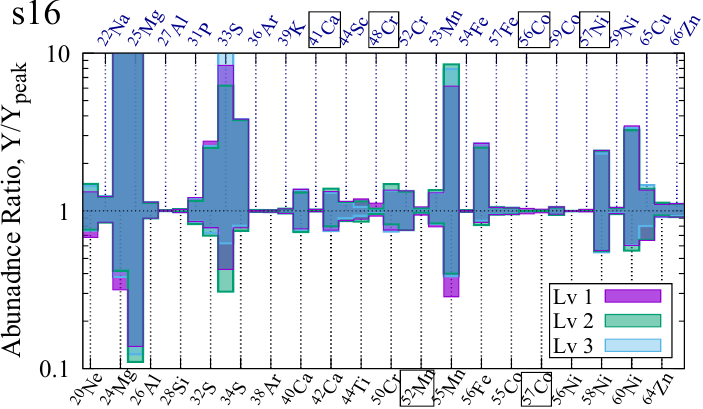}
\vspace{0.005\vsize}

\includegraphics[width=1\hsize]{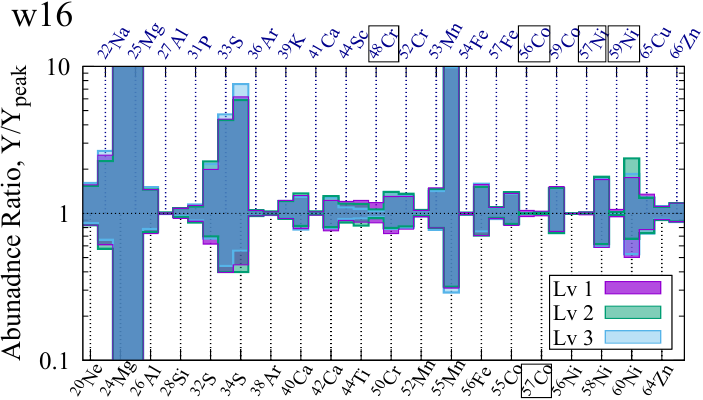}
\vspace{0.005\vsize}

\includegraphics[width=1\hsize]{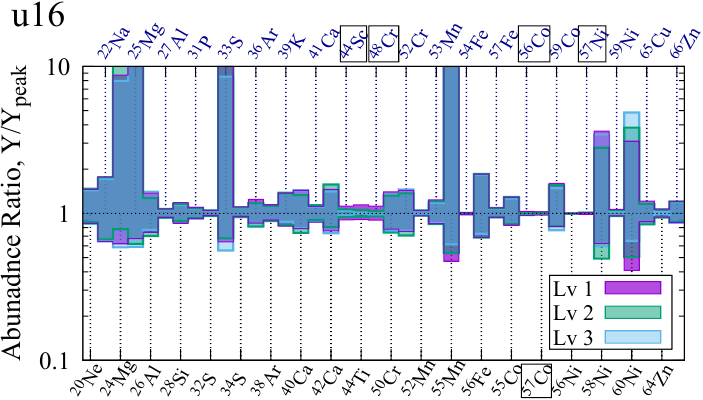}

\caption{Uncertainty ranges (90\% intervals around the peak value $y_{\mathrm{peak}}$) obtained from the MC variations for explosion models: s16 (top), w16 (middle), and u16 (bottom), shown for Level~1–3 MC runs. Nuclei marked with squares indicate key products listed in Tables~\ref{tab:key_s16}, \ref{tab:key_w16}, and \ref{tab:key_u16}. Selected nuclides are $\iso{Ne}{20}$, $\iso{Na}{22}$, $\iso{Mg}{24, 25}$, $\iso{Al}{26, 27}$, $\iso{Si}{28}$, $\iso{P}{31}$, $\iso{S}{32,33,34}$, $\iso{Ar}{36, 38}$, $\iso{K}{39}$, $\iso{Ca}{40,41,42}$, $\iso{Sc}{44}$, $\iso{Ti}{44}$, $\iso{Cr}{48, 50, 52}$, $\iso{Mn}{52, 53, 55}$, $\iso{Fe}{54, 56, 57}$, $\iso{Co}{55, 56, 57, 59}$, $\iso{Ni}{56, 57, 58, 59, 60}$, $\iso{Cu}{65}$, and $\iso{Zn}{64, 66}$, arranged in this order.}
\label{fig:y_unc}
\end{figure}

\begin{figure}

\centering
\includegraphics[width=\hsize]{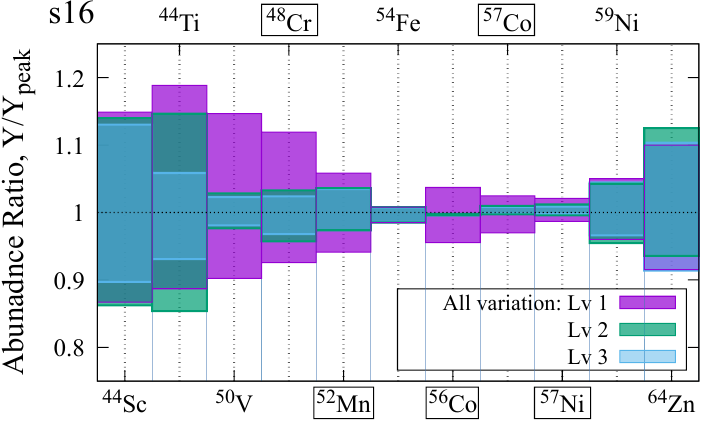}
\vspace{0.005\vsize}

\includegraphics[width=\hsize]{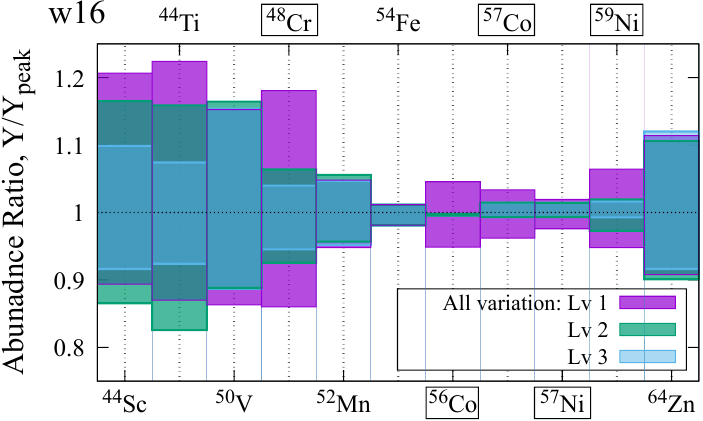}
\vspace{0.005\vsize}

\includegraphics[width=\hsize]{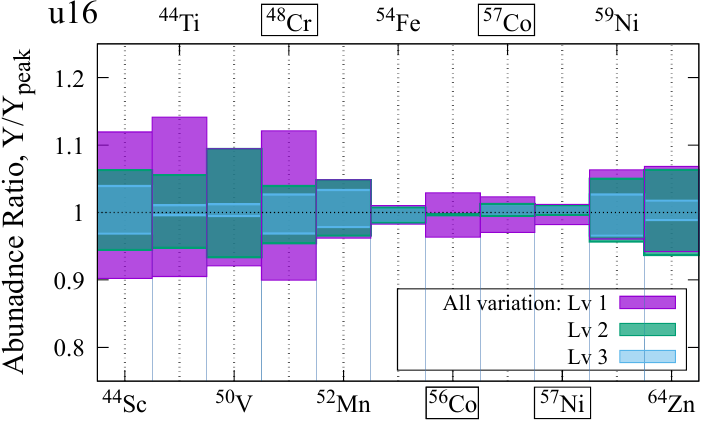}

\caption{Uncertainty ranges obtained from the MC variations for s16 (left), w16 (middle), and u16 (right). Same as Figure~\ref{fig:y_unc}, but for selected nuclei.}
\label{fig:y_unc_selec}
\end{figure}

\begin{table*}
\caption{Key reactions and decays for the s16 model. The underlined nuclides are stable ones (beyond p and $\alpha$), all others are radioactive. The notation used for nuclear reactions (pairs) and weak decays is explained in Sec.~\ref{sec:mcmethod}.}
\label{tab:key_s16}
\begin{tabular}{lccccccccccccccc}
\hline
Nuclide & $r_{{\rm corr},0}$ & Key rate &$r_{{\rm corr},1}$ & Key rate &
$r_{{\rm corr},2}$ & Key rate & $X_0$ & Compound nucleus\\
&&Level 1&&Level 2&&Level 3 & (3, 5 GK)\\
\hline
$\iso{Ca}{41}$ & $-0.67$ & $\underline{\iso{Ar}{38}} + \alpha \leftrightarrow {\rm n} + \iso{Ca}{41}$ &&&&&1.00, 0.94&$\underline{\iso{Ca}{42}}$\\
$\iso{Cr}{48}$ & $-0.82$ & $\iso{Cr}{48} + \alpha \leftrightarrow {\rm p} + \underline{\iso{Mn}{51}}$ &&&&&0.78, 0.50&$\iso{Fe}{52}$\\
&&& $0.65$ & $\iso{Ti}{44} + \alpha \leftrightarrow {\rm p} + \underline{\iso{V}{47}}$ &&&0.28, 0.14&$\iso{Cr}{48}$\\
&&&&& $0.88$ & $\iso{Ca}{40} + \alpha \leftrightarrow \gamma + \iso{Ti}{44}$ &1.00, 1.00&$\iso{Ti}{44}$\\
$\iso{Mn}{52}$ & $-0.70$ & $\iso{Fe}{52} + \alpha \leftrightarrow {\rm p} + \iso{Co}{55}$ &&&&&0.84, 0.57&$\iso{Ni}{56}$\\
&&&&& $0.67$ & ${}^{52}{\rm{Fe}}(\varepsilon+\beta^+){}^{52}{\rm{Mn}}$ &&--\\
$\iso{Co}{56}$ & $1.00$ & ${}^{56}{\rm{Ni}}(\varepsilon+\beta^+){}^{56}{\rm{Co}}$ &&&&&&--\\
&&& $-0.69$ & $\iso{Ni}{56} + \alpha \leftrightarrow {\rm p} + \iso{Cu}{59}$ &&&0.74, 0.46&\underline{$\iso{Zn}{60}$}\\
${}^{57}{\rm Co}$ & $0.92$ & ${}^{57}{\rm{Ni}}(\varepsilon+\beta^+){}^{57}{\rm{Co}}$ &&&&&&--\\
&&& $-0.65$ & $\underline{\iso{Ne}{20}} + \alpha \leftrightarrow \gamma + \underline{\iso{Mg}{24}}$ &&&1.00, 0.99&\underline{$\iso{Mg}{24}$}\\
$\iso{Ni}{57}$ & $-0.83$ & $\iso{Co}{57} + {\rm p} \leftrightarrow {\rm n} + \iso{Ni}{57}$ &&&&&0.89, 0.68&\underline{$\iso{Ni}{58}$}\\
\hline
\end{tabular}
\end{table*}

\begin{table*}
\caption{Key reactions and decays for the w16 model. Same as Table~\ref{tab:key_s16}.}
\label{tab:key_w16}
\begin{tabular}{lccccccccccccccc}
\hline
Nuclide & $r_{{\rm corr},0}$ & Key rate &$r_{{\rm corr},1}$ & Key rate &
$r_{{\rm corr},2}$ & Key rate & $X_0$ & Compound nucleus\\
&&Level 1&&Level 2&&Level 3 & (3, 5 GK)\\
\hline
$\iso{Cr}{48}$ & $-0.81$ & $\iso{Cr}{48} + \alpha \leftrightarrow {\rm p} + \underline{\iso{Mn}{51}}$ &&&&&0.78, 0.50&$\iso{Fe}{52}$\\
&&& $0.65$ & $\iso{Ti}{44} + \alpha \leftrightarrow {\rm p} + \underline{\iso{V}{47}}$ &&&0.28, 0.14&$\iso{Cr}{48}$\\
&&&&& $0.89$ & $\underline{\iso{Ca}{40}} + \alpha \leftrightarrow \gamma + \iso{Ti}{44}$ &1.00, 1.00&$\iso{Ti}{44}$\\
$\iso{Co}{56}$ & $1.00$ & ${}^{56}{\rm{Ni}}(\varepsilon+\beta^+){}^{56}{\rm{Co}}$ &&&&&&--\\
&&& $-0.65$ & ${}^{56}{\rm{Ni}} + \alpha \leftrightarrow {\rm p} + {}^{59}{\rm{Cu}}$ &&&0.74, 0.46&\underline{$\iso{Zn}{60}$}\\
$\iso{Co}{57}$ & $0.91$ & ${}^{57}{\rm{Ni}}(\varepsilon+\beta^+){}^{57}{\rm{Co}}$ &&&&&&--\\
$\iso{Ni}{57}$ & $-0.85$ & ${}^{57}{\rm{Co}} + {\rm p} \leftrightarrow {\rm n} + {}^{57}{\rm{Ni}}$ &&&&&0.89, 0.68&\underline{$\iso{Ni}{58}$}\\
$\iso{Ni}{59}$ & $-0.85$ & ${}^{59}{\rm{Cu}} + {\rm p} \leftrightarrow \gamma + \underline{\iso{Zn}{60}}$ &&&&&0.97, 0.87&\underline{$\iso{Zn}{60}$}\\
\hline
\end{tabular}
\end{table*}

\begin{table*}
\caption{Key reactions and decays for the u16 model. Same as Table~\ref{tab:key_s16}. Only $\iso{Sc}{44}$ decreases in abundance after the NSE.}
\label{tab:key_u16}
\begin{tabular}{lccccccccccccccc}
\hline
Nuclide & $r_{{\rm corr},0}$ & Key rate &$r_{{\rm corr},1}$ & Key rate &
$r_{{\rm corr},2}$ & Key rate & $X_0$ & Compound nucleus\\
&&Level 1&&Level 2&&Level 3 & (3, 5 GK)\\
\hline
${}^{44}{\rm Sc}$ & $-0.72$ & ${}^{44}{\rm{Ti}} + \alpha \leftrightarrow {\rm p} + \underline{{}^{47}{\rm{V}}}$ &&&&&0.28, 0.14&$\iso{Cr}{48}$\\
&&& $0.74$ & $\underline{{}^{40}{\rm{Ca}}} + \alpha \leftrightarrow \gamma + {}^{44}{\rm{Ti}}$ &&&1.00, 1.00&$\iso{Ti}{44}$\\
&&&&& $0.83$ & ${}^{44}{\rm{Ti}}(\varepsilon){}^{44}{\rm{Sc}}$ &&--\\
${}^{48}{\rm Cr}$ & $-0.75$ & ${}^{48}{\rm{Cr}} + \alpha \leftrightarrow {\rm p} + \underline{{}^{51}{\rm{Mn}}}$ &&&&&0.78, 0.50&$\iso{Fe}{52}$\\
&&&&& $-0.84$ & ${}^{49}{\rm{Mn}} + {\rm p} \leftrightarrow \gamma + \underline{{}^{50}{\rm{Fe}}}$ &0.78, 0.67&\underline{$\iso{Fe}{50}$}\\
&&& $-0.66$ & ${}^{52}{\rm{Fe}} + \alpha \leftrightarrow {\rm p} + {}^{55}{\rm{Co}}$ &&&0.84, 0.57&$\iso{Ni}{56}$\\
${}^{56}{\rm Co}$ & $1.00$ & ${}^{56}{\rm{Ni}}(\varepsilon+\beta^+){}^{56}{\rm{Co}}$ &&&&&&--\\
&&&&& $-0.71$ & ${}^{56}{\rm{Ni}} + \alpha \leftrightarrow {\rm p} + {}^{59}{\rm{Cu}}$ &0.74, 0.46&\underline{$\iso{Zn}{60}$}\\
${}^{57}{\rm Co}$ & $0.87$ & ${}^{57}{\rm{Ni}}(\varepsilon+\beta^+){}^{57}{\rm{Co}}$ &&&&&&--\\
${}^{57}{\rm Ni}$ & $-0.79$ & ${}^{57}{\rm{Co}} + {\rm p} \leftrightarrow {\rm n} + {}^{57}{\rm{Ni}}$ &&&&&0.89, 0.68&\underline{$\iso{Ni}{58}$}\\
&&& $-0.66$ & ${}^{59}{\rm{Cu}} + {\rm p} \leftrightarrow \gamma + \underline{{}^{60}{\rm{Zn}}}$ &&&0.97, 0.87&\underline{$\iso{Zn}{60}$}\\
\hline
\end{tabular}
\end{table*}

\begin{figure*}
\centering
\includegraphics[width=0.32\hsize]{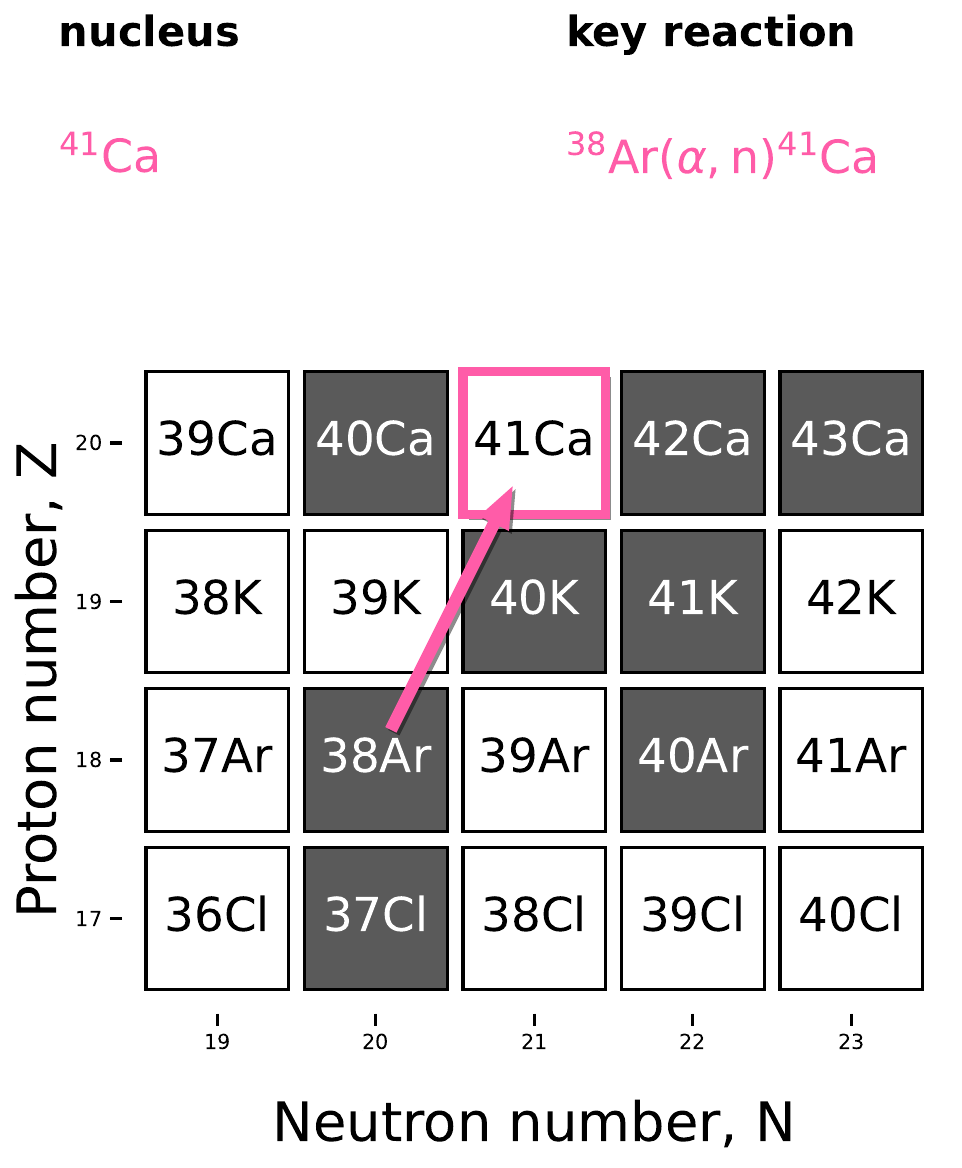}
\hspace{0.02\hsize}
\includegraphics[width=0.65\hsize]{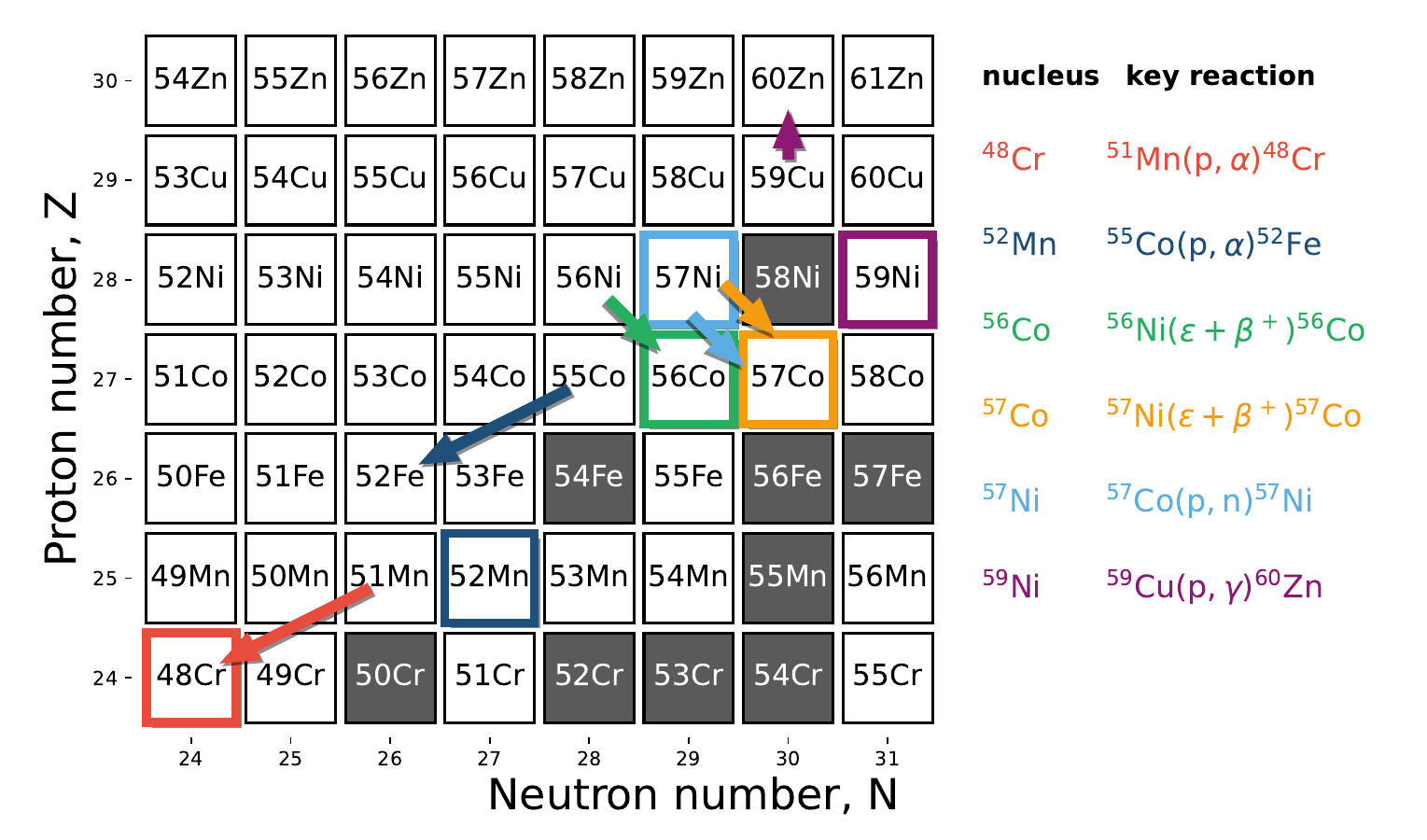}
\caption{Key reactions and corresponding nucleus on the $N$--$Z$ plane. Level~1 rates listed in Tables~\ref{tab:key_s16}, \ref{tab:key_w16}, and \ref{tab:key_u16} are adopted. Key reactions and decays are plotted in the same color as the corresponding nucleus.}
\label{fig:key_flow}
\end{figure*}

The uncertainty range of the nucleosynthesis products, defined as the 90\% interval around the peak value $y_{\mathrm{peak}}$ and resulting from variations in reaction and decay rates, is shown as the normalized abundance $Y / Y_{\mathrm{peak}}$ in Figure~\ref{fig:y_unc}. The results for the s16 and w16 models are generally consistent, exhibiting a similarly large range of uncertainty for products with significant production. In contrast, the results for the u16 model differ markedly from these due to the much lower metallicity of the progenitor. Some nuclides exhibited abundance uncertainties exceeding $10$--$100$, for example $^{24,25}$Mg and $^{33,34}$S, and $2$--$5$ for a few Fe-peak nuclei. These nuclides are produced only in very small amounts during explosive nucleosynthesis, resulting in large relative deviations.

{In Figure~\ref{fig:y_unc_selec}, we ignore such spurious values for nuclides with final abundances of $Y \lesssim 10^{-5}$, since even large relative enhancements remain negligible and no important reactions are excluded. According to the ``Level 1'' results in Figure~\ref{fig:y_unc_selec}, which include all rate variations defined in Table~\ref{tab:unc}, the uncertainty range spans from a few percent to about 20\% and is not significantly asymmetric between the lower and upper limits. Fe and Ni isotopes, which are mostly produced under NSE conditions, show smaller uncertainties, whereas isotopes with $A < 50$ exhibit relatively large uncertainties. The latter nuclei are produced in the incomplete Si-burning layer, so their uncertainties are reasonably attributed to several nuclear reactions whose rates are not well determined.

Furthermore, we conducted a statistical analysis of the MC nucleosynthesis results and calculated correlation coefficients to identify key reactions as explained in Section~\ref{sec:mcmethod}. Tables~\ref{tab:key_s16}, \ref{tab:key_w16}, and \ref{tab:key_u16} summarize the key reactions found for each model along with their correlation coefficients. The bidirectional arrow in the reaction notation indicates that both forward and reverse reactions were varied simultaneously because they are connected by detailed balance.

As in our previous work, we investigate several ``levels'' of key reactions. Level-1 reactions (with correlation $r_\mathrm{corr,0}$) are the most important reactions. Level-2 reactions (with correlation $r_\mathrm{corr,1}$) are identified by repeating the MC procedure without variation of the level-1 reactions. Thus, level 2 reactions only become important once level-1 reactions have been determined accurately. Likewise, level-3 reactions are obtained without variation of level-1 and level-2 rates. The normalized abundances obtained at the different levels also are shown in Figures~\ref{fig:y_unc} and \ref{fig:y_unc_selec}. Note that although the overall uncertainty reduces with every level, the resulting uncertainty bar may be shifted relative to the previous levels. It is especially clear from Figure~\ref{fig:y_unc_selec} that the uncertainty range becomes smaller when the uncertainty of key reactions is removed. In general, the reduction from Level~1 to Level~2 is significant. Although $\iso{Ti}{44}$ does not have its own key reactions, the uncertainty is still considerably reduced. This is because the $\iso{Ti}{44}$ abundance is influenced by neighboring reactions, some of which involve key nuclei such as $\iso{Cr}{48}$ (see Section~\ref{res:mc-layer}).

The first column in Tables~\ref{tab:key_s16}, \ref{tab:key_w16}, and \ref{tab:key_u16} specifies the nuclide of which the abundance is significantly affected by the shown key rate. Also shown in the tables are the ground-state contributions $X_0$ to the stellar reaction rate at two temperatures, 3 GK and 5 GK. The values are taken from \cite{2012ApJS..201...26R}. They are intended to guide experimenters as they indicate what percentage of the stellar rate can be constrained by a measurement of a reaction on the ground state of the target nucleus. The values are given for the reaction direction with positive $Q$-value ($^{41}$Ca(n,$\alpha$)$^{38}$Ar, $^{47}$V(p,$\alpha$)$^{44}$Ti, $^{59}$Cu(p,$\alpha$)$^{56}$Ni, $^{57}$Ni(n,p)$^{57}$Co) and for captures regardless of $Q$-value as their $X_0$ values are larger than for their reverse reactions.

The final abundances of the key nuclei listed in Tables~\ref{tab:key_s16}–\ref{tab:key_u16} generally increase due to explosive nucleosynthesis, i.e., they are produced after NSE, with the exception of $\iso{Sc}{44}$ in the u16 model (Table~\ref{tab:key_u16}), which is destroyed after NSE. In this case, explosive oxygen burning is insufficient to raise the abundance of $\iso{Sc}{44}$ above its initial value in the pre-collapse NSE core. Its key reaction, $\iso{Ti}{44}(\alpha,\iso{p}{})\iso{V}{47}$, has also been discussed in previous studies, but in a different context, for example as one of the reactions affecting $^{44}$Ti \citep{2000PhRvL..84.1651S, 2020PhRvC.102c5806C}. Here, we also identify this reaction as a level-2 key reaction for the production of $^{48}$Cr in models s16 and w16.

In \cite{2012ApJ...750...18A} the reactions $^{59}$Cu(p,$\gamma$) and $^{59}$Cu(p,$\alpha$) have been found to be important in determining whether the $\nu p$-process path continues beyond mass $A=59$. Our present investigation does not include the $\nu p$-process but these reactions appear as key reactions of levels 1, 2, and 3 for Co and Ni isotopes. The proton capture channel appears as level-1 key reaction for $\iso{Ni}{59}$ in model w16 and as level-2 key reaction for $\iso{Ni}{57}$ in model u16. The (p,$\alpha$) reaction shows up as being important for production of $\iso{Co}{56}$ in all models, as level-2 key reaction in models s16 and w16, and as level-3 key reaction in model u16.

Figure~\ref{fig:key_flow} shows the combinations of key reactions and their corresponding nuclides on a section of the nuclear chart. When $\beta$ decay (or electron capture) is the key reaction, it serves as a key reaction for the parent or daughter nucleus (see the first column in Tables~\ref{tab:key_s16}--\ref{tab:key_u16}). In contrast, for other nuclear reactions, a key reaction may also play an indirect role in the production of other nuclides, not directly involved in the reaction. This is due to the fact that in nucleosynthesis, where multiple nuclear reactions are involved, the impact of uncertainties also appears in a complex manner. Parts of 5th paragraph in the right of page~8 was revised. These results demonstrate the usefulness and necessity of the MC approach in nucleosynthesis sensitivity studies. Without the MC approach, one must either individually investigate the sensitivity of all reaction rates, significantly increasing the computational cost, or narrow down the surrounding reactions, with the risk of overlooking key rates. Once the key reaction rates are identified through MC analysis, regardless of the specific reactions or decays involved, they can be subjected to more detailed studies focusing on individual rates.

\begin{figure*}
\centering

\includegraphics[width=0.47\hsize]{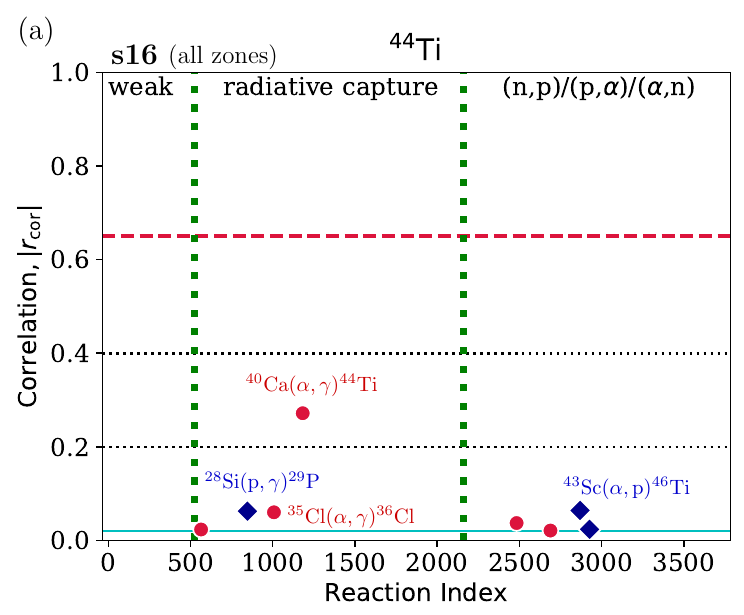}
\hspace{0.05\hsize}
\includegraphics[width=0.47\hsize]{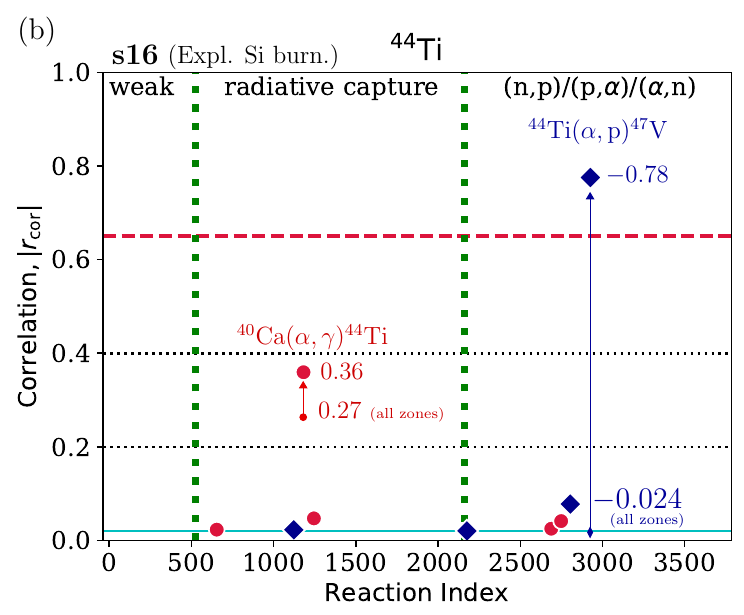}

\caption{Correlation factors of $\iso{Ti}{44}$ (s16 model) production with respect to all varied reaction and decay rates, labeled by their ``reaction index'' numbers. The plot shows correlations with $|r_{\mathrm{cor}}| \ge 0.02$: positive correlations are indicated by red circles, while negative correlations are shown by blue diamonds. The red dashed line corresponds to the threshold for the key reaction, $0.65$. (a) Results of the analysis including all stellar zones. (b) Results of the analysis restricted to the Si-burning layers.}
\label{fig:cor_sib_ti44}

\end{figure*}


\begin{table}
\centering
\caption{Key nuclear reactions (level 1) from MC calculations with fixed weak rates.}
\label{tab:key_s16aa}
\begin{tabular}{lcccccccccc}
\hline
Nuclide  & Key rate & s16 & w16 & u16 \\
&& $r_{{\rm corr},1}$ & $r_{{\rm corr},1}$ & $r_{{\rm corr},1}$ \\
\hline
$\iso{Ca}{41}$ & $\iso{Ar}{38} + \alpha \leftrightarrow \rm{n} + \iso{Ca}{41}$ & $-0.67$ \\
$\iso{Sc}{44}$ & $\iso{Ti}{44} + \alpha \leftrightarrow \rm{p} + \iso{V}{47}$  & & & $-0.76$ \\
$\iso{Cr}{48}$ & $\iso{Cr}{48} + \alpha \leftrightarrow \rm{p} + \iso{Mn}{51}$ & $-0.82$ & $-0.81$ &$-0.75$ \\
$\iso{Mn}{52}$ & $\iso{Fe}{52} + \alpha \leftrightarrow \rm{p} + \iso{Co}{55}$ & $-0.73$ &&$-0.66$ \\
$\iso{Co}{57}$ & $\iso{Co}{57} + \rm{p} \leftrightarrow \rm{n} + \iso{Ni}{57}$ & $-0.85$ & $-0.85$ & $-0.80$ \\
$\iso{Ni}{59}$ & $\iso{Cu}{59} + \rm{p} \leftrightarrow \gamma + \iso{Zn}{60}$ & & $-0.85$ \\
\hline
\end{tabular}
\end{table}

As uncertainties in decay half-lives may mask other important reactions, we also performed MC runs without including decays in the variation for the three progenitor models. The resulting level-1 key reactions are shown in Table~\ref{tab:key_s16aa}. Removing the decay half-lives from the MC variation led to the appearance of the level-1 key reaction $\iso{Co}{57} + \rm{p} \leftrightarrow \rm{n} + \iso{Ni}{57}$ instead of the $^{57}$Ni decay in all three models. There was no key reaction affecting $^{56}$Co anymore without variation of the decays. All other level-1 key reactions remained unchanged.

\subsection{MC analysis of complete silicon burning}
\label{res:mc-layer}

We also performed statistical analyses for individual layers (e.g., for the explosive Si-burning layer; see Table~\ref{tab:models}), applying the same MC results, weighting factors, and methodology, but restricting the analysis to different zone ranges. We find that the values of $|r_{\rm cor}|$ are enhanced for some rates, and few reactions or decays exhibit $|r_{\rm cor}| \ge 0.65$, meeting our threshold for designation as key reactions. For the complete Si-burning layer, one additional rate, $\iso{Ti}{44}(\alpha,{\rm p})\iso{V}{47}$, satisfies the criterion for $\iso{Ti}{44}$ production across all explosion models. The resulting correlation factors, $r_{\mathrm{cor}}$, for $\iso{Ti}{44}$ in the s16 model are shown in Figure~\ref{fig:cor_sib_ti44} among varied reactions and decays\footnote{We varied a total of $\sim 8,000$ reactions, but the number of index entries is about half of that, because forward and reverse reactions were not varied independently.}. Selected $|r_\mathrm{cor}|$ are plotted for the case of all zones (Figure~\ref{fig:cor_sib_ti44}a and for the explosive Si-burning layer (Figure~\ref{fig:cor_sib_ti44}b). We identify an additional key reaction, $\iso{Ti}{44}(\alpha,\mathrm{p})\iso{V}{47}$, and another reaction, $\iso{Ca}{40}(\alpha,\gamma)\iso{Ti}{44}$, both of which significantly increase $r_{\mathrm{cor}}$. These reactions are already included among the key reactions responsible for the production of other nuclei, such as $\iso{Sc}{44}$ and $\iso{Cr}{48}$, in Levels~1 to 3.

The correlations of $\iso{Ti}{44}(\alpha,\mathrm{p})\iso{V}{47}$ (from $r_{\mathrm{cor}} = -0.024$ to $-0.78$) and $\iso{Ca}{40}(\alpha,\gamma)\iso{Ti}{44}$ (from $r_{\mathrm{cor}} = 0.27$ to $0.36$) become significantly stronger when only explosive Si-burning zones are included in the analysis. Similar trends are found for the other models. In the explosive Si-burning layers, $r_{\mathrm{cor}} = -0.76$ and $-0.75$ for $\iso{Ti}{44}(\alpha,\mathrm{p})\iso{V}{47}$ in the w16 and u16 models, respectively, and $r_{\mathrm{cor}} = 0.37$ and $0.40$ for $\iso{Ca}{40}(\alpha,\gamma)\iso{Ti}{44}$ in the u16 and w16 models, respectively. According to our current criterion, $\iso{Ca}{40}(\alpha,\gamma)\iso{Ti}{44}$ is not identified as a key reaction; however, it becomes the dominant reaction in the explosive Ne-burning layers ($M - M_{\mathrm{cut}} = 0.2$–$0.3M_{\odot}$) of the s16 model (Figure~\ref{fig:std_nucl}). Although its impact is reduced when averaged over the entire star, it may still be observationally significant if the contribution from this particular ejecta layer can be isolated or weighted by a specific observational probe.

In addition to $\iso{Ti}{44}(\alpha,{\rm p})\iso{V}{47}$, additional decay contributing to $\iso{V}{48}$ production is identified only in the u16 model: $\iso{Cr}{48}(\varepsilon+\beta^+)\iso{V}{48}$ ($r_\mathrm{cor}=0.90$), whose ground state half-life is $21.56~\mathrm{h}$. The reaction $\iso{Ti}{44}(\alpha,\mathrm{p})\iso{V}{47}$, relevant to the production of $\iso{Sc}{44}$, also appears as a key reaction ($r_\mathrm{cor} = 0.90$) in w16, although in our previous full-layer analysis it was identified only for u16. This reaction is already identified as key for several nuclei with $A \lesssim 50$. Note that our nucleosynthesis calculations are terminated after $10^4$~s ($2.78$~h), so the inferred importance of key nuclei for astronomical observations depends sensitively on the time after explosion.

\subsection{Impacts on Supernova observables}
\label{sec:obs}

Several key radioactive nuclei produced in explosive nucleosynthesis have been investigated in connection with optical observations. \cite{2019BAAS...51c...2T} identified radioactive SN products which potentially could be observed in SN remnants by their decay lines. In addition to the well known cases of $\iso{Al}{26}$, $\iso{Ti}{44}$, and $\iso{Ni}{56}$, these are $\iso{K}{43}$, $\iso{Sc}{47}$, $\iso{V}{48}$, $\iso{Ca}{47}$, $\iso{Sc}{44}$, $\iso{Cr}{48,51}$, $\iso{Mn}{52}$, $\iso{Fe}{59}$, $\iso{Co}{56,57}$, and $\iso{Ni}{57}$. Concerning $\iso{Al}{26}$, this Al isotope is produced in several distinct phases during stellar evolution and also explosively \citep[see, e.g.,][]{2025ApJ...991...21F}. An analysis of key reactions for this nuclide is not within the scope of the present work.

Inspection of Tables~\ref{tab:key_s16}$-$\ref{tab:key_u16} shows that key reactions for the explosive production of $\iso{Sc}{44}$, $\iso{Cr}{48}$, $\iso{Mn}{52}$, $\iso{Co}{56,57}$, and $\iso{Ni}{57}$ were found in our approach. The nuclide $\iso{Ni}{56}$ is a typical isotope produced by explosive nucleosynthesis and one of the most important nuclei involved in the observation of supernova light curves. However, its production corresponds to the peak of the nuclear distribution in the vicinity of the NSE and is not driven by individual nuclear reactions. Therefore, it is not surprising that individual nuclear reaction rates cannot be identified as being important for the production of Fe and Ni isotopes. The weak decay $\iso{Ni}{56} \rightarrow \iso{Co}{56}$, which is a well-known SN light-curve energy source, can be identified as the dominant process affecting the abundance of $\iso{Ni}{56}$ about $10~\mathrm{h}$ after the explosion. On the other hand, we find that the reaction $\iso{Co}{57}(\mathrm{p},\mathrm{n})\iso{Ni}{57}$ is a key process for $\iso{Ni}{57}$ production, and $\iso{Ni}{57}$ is also an observable nucleus, similar to other Fe and Ni isotopes, particularly in terms of its ratio to $\iso{Ni}{56}$ \citep[see, e.g.,][]{1996ApJ...460..408T, 1997ApJ...486.1026N}. Within our approach of studying explosive nucleosynthesis by post-processing existing progenitor models we only find a key reaction for $\iso{Ti}{44}$ when focusing on the region of complete Si burning (see Section~\ref{res:mc-layer}).

The production of $\iso{Ni}{56}$ and $\iso{Ti}{44}$ also strongly depends on the explosion mechanism and thus on the mass cut. In the PUSH method, the mass cut is obtained self-consistently and is not a free parameter. The dependence on the explosion mechanism of the production of $\iso{Ti}{44}$ has been investigated in previous studies \citep[e.g.,][]{2010ApJS..191...66M, 2020ApJ...890...35A}. \citet{2024ApJ...974...39W} recently pointed out that neutrino-driven winds in long-term multi-D SN simulations can yield additional production of $\iso{Ti}{44}$. Additionally, \cite{2020ApJ...898....5S} found additional reactions impacting the production of $\iso{Ti}{44}$ and $\iso{Ni}{56}$ when also taking into account uncertainties in the stellar evolution phase.

\section{Summary and conclusion}
\label{sec:conclusion}

We have investigated nuclear uncertainties in iron-group nuclides (up to $A=80$) in explosive nucleosynthesis of core-collapse supernovae within a Monte Carlo approach, in which we simultaneously varied about 8,000 reactions. Three explosion models based on a $16 M_\odot$ progenitor with different metallicities were post-processed using several hundred explosive trajectories each, covering the innermost ejected layers of the stars, obtained using the PUSH method.

We obtained abundance uncertainties stemming from nuclear input, i.e., from uncertainties in reaction rates and decay half-lives. Key reactions were identified for a number of nuclides, whereas for the majority of nuclides in the full reaction network no single key reaction was found. This implies that either their abundances are insensitive to reaction rate variations or several reactions contribute more or less equally to their abundances and abundance uncertainties. In the latter case, only general improvements in reaction rate predictions can reduce the remaining uncertainty because an experimental determination of such a large range of reaction rates seems unlikely. On the other hand, in all of the identified key reactions at least one of the target or final nuclei is stable or the compound nucleus is stable. Furthermore, with the exception of $\iso{Ti}{44} + \alpha \leftrightarrow \rm{p} + \iso{V}{47}$, the ground-state contributions to the stellar rate are significant. These facts are promising for the feasibility of experimentally reducing the nuclear uncertainties in the astrophysical production in this nuclear mass range.

We note that the results and conclusions of the present study are influenced by physical uncertainties in the explosion model, including those originating from progenitor evolution. Nevertheless, as with discussions of nucleosynthesis in one-dimensional supernova models, if the present study is regarded as a phenomenological approach to reproducing observables, the nucleosynthesis process should be understood as capturing the overall framework, similar to studies employing the ``PUSH'' method. Accordingly, in representative scenarios, the impact of uncertainties in nuclear reaction rates should be appropriately reflected. However, in realistic explosions that incorporate multidimensional effects, the quantitative aspects presented here may require revision, warranting confirmation in future, more advanced investigations.

\section*{Acknowledgments}

The authors acknowledge C.~Evans, T.~Furuno, and R.~Sawada for fruitful discussions. They also thank the anonymous referee for a careful reading and for providing useful comments. The project was financially supported by JSPS KAKENHI (JP20H05648, JP21H01087, JP24H00008, JP25H01273) and by the US National Science Foundation under Grant No. OISE-1927130 (IReNA). NN was supported by the RIKEN Incentive Research Project (FY2024--2025). Parts of the computations were carried out on facilities at CfCA, NAOJ and YITP, Kyoto University. The work at NC State was supported by the United States Department of Energy, Office of Science, Office of Nuclear Physics (award number DE-FG02-02ER41216).

\section*{DATA AVAILABILITY}

All data for the correlation factors $r_\mathrm{cor}$ of key nuclei in each model are publicly available on GitHub at \url{https://github.com/nnobuya/mc_ccsn_push}. Numerical tables and figures, such as Figure~\ref{fig:cor_sib_ti44}, will be made available after publication. The data before the public release or other relevant data are available from the corresponding author (N.N.) upon reasonable request.




\bibliographystyle{mnras}
\bibliography{ref} 

\begin{thebibliography}{}
\makeatletter
\relax
\def\mn@urlcharsother{\let\do\@makeother \do\$\do\&\do\#\do\^\do\_\do\%\do\~}
\def\mn@doi{\begingroup\mn@urlcharsother \@ifnextchar [ {\mn@doi@}
  {\mn@doi@[]}}
\def\mn@doi@[#1]#2{\def\@tempa{#1}\ifx\@tempa\@empty \href
  {http://dx.doi.org/#2} {doi:#2}\else \href {http://dx.doi.org/#2} {#1}\fi
  \endgroup}
\def\mn@eprint#1#2{\mn@eprint@#1:#2::\@nil}
\def\mn@eprint@arXiv#1{\href {http://arxiv.org/abs/#1} {{\tt arXiv:#1}}}
\def\mn@eprint@dblp#1{\href {http://dblp.uni-trier.de/rec/bibtex/#1.xml}
  {dblp:#1}}
\def\mn@eprint@#1:#2:#3:#4\@nil{\def\@tempa {#1}\def\@tempb {#2}\def\@tempc
  {#3}\ifx \@tempc \@empty \let \@tempc \@tempb \let \@tempb \@tempa \fi \ifx
  \@tempb \@empty \def\@tempb {arXiv}\fi \@ifundefined
  {mn@eprint@\@tempb}{\@tempb:\@tempc}{\expandafter \expandafter \csname
  mn@eprint@\@tempb\endcsname \expandafter{\@tempc}}}

\bibitem[\protect\citeauthoryear{{Andrews}, {Fryer}, {Even}, {Jones}  \&
  {Pignatari}}{{Andrews} et~al.}{2020}]{2020ApJ...890...35A}
{Andrews} S.,  {Fryer} C.,  {Even} W.,  {Jones} S.,   {Pignatari} M.,  2020,
  \mn@doi [\apj] {10.3847/1538-4357/ab64f8}, \href
  {https://ui.adsabs.harvard.edu/abs/2020ApJ...890...35A} {890, 35}

\bibitem[\protect\citeauthoryear{{Arcones}, {Fr{\"o}hlich}  \&
  {Mart{\'\i}nez-Pinedo}}{{Arcones} et~al.}{2012}]{2012ApJ...750...18A}
{Arcones} A.,  {Fr{\"o}hlich} C.,   {Mart{\'\i}nez-Pinedo} G.,  2012, \mn@doi
  [\apj] {10.1088/0004-637X/750/1/18}, \href
  {https://ui.adsabs.harvard.edu/abs/2012ApJ...750...18A} {750, 18}

\bibitem[\protect\citeauthoryear{{Boggs} et~al.,}{{Boggs}
  et~al.}{2015}]{2015Sci...348..670B}
{Boggs} S.~E.,  et~al., 2015, \mn@doi [Science] {10.1126/science.aaa2259},
  \href {https://ui.adsabs.harvard.edu/abs/2015Sci...348..670B} {348, 670}

\bibitem[\protect\citeauthoryear{{Burrows}}{{Burrows}}{2013}]{2013RvMP...85..245B}
{Burrows} A.,  2013, \mn@doi [Reviews of Modern Physics]
  {10.1103/RevModPhys.85.245}, \href
  {https://ui.adsabs.harvard.edu/abs/2013RvMP...85..245B} {85, 245}

\bibitem[\protect\citeauthoryear{{Cescutti}, {Hirschi}, {Nishimura}, {Hartogh},
  {Rauscher}, {Murphy}  \& {Cristallo}}{{Cescutti}
  et~al.}{2018}]{2018MNRAS.478.4101C}
{Cescutti} G.,  {Hirschi} R.,  {Nishimura} N.,  {Hartogh} J.~W.~d.,  {Rauscher}
  T.,  {Murphy} A. S.~J.,   {Cristallo} S.,  2018, \mn@doi [\mnras]
  {10.1093/mnras/sty1185}, \href
  {https://ui.adsabs.harvard.edu/abs/2018MNRAS.478.4101C} {478, 4101}

\bibitem[\protect\citeauthoryear{{Chipps}, {Adsley}, {Couder}, {Hix}, {Meisel}
  \& {Schmidt}}{{Chipps} et~al.}{2020}]{2020PhRvC.102c5806C}
{Chipps} K.~A.,  {Adsley} P.,  {Couder} M.,  {Hix} W.~R.,  {Meisel} Z.,
  {Schmidt} K.,  2020, \mn@doi [\prc] {10.1103/PhysRevC.102.035806}, \href
  {https://ui.adsabs.harvard.edu/abs/2020PhRvC.102c5806C} {102, 035806}

\bibitem[\protect\citeauthoryear{{Curtis}, {Ebinger}, {Fr{\"o}hlich}, {Hempel},
  {Perego}, {Liebend{\"o}rfer}  \& {Thielemann}}{{Curtis}
  et~al.}{2019}]{2019ApJ...870....2C}
{Curtis} S.,  {Ebinger} K.,  {Fr{\"o}hlich} C.,  {Hempel} M.,  {Perego} A.,
  {Liebend{\"o}rfer} M.,   {Thielemann} F.-K.,  2019, \mn@doi [\apj]
  {10.3847/1538-4357/aae7d2}, \href
  {https://ui.adsabs.harvard.edu/abs/2019ApJ...870....2C} {870, 2}

\bibitem[\protect\citeauthoryear{{Cyburt} et~al.,}{{Cyburt}
  et~al.}{2010}]{2010ApJS..189..240C}
{Cyburt} R.~H.,  et~al., 2010, \mn@doi [\apjs] {10.1088/0067-0049/189/1/240},
  \href {https://ui.adsabs.harvard.edu/abs/2010ApJS..189..240C} {189, 240}

\bibitem[\protect\citeauthoryear{{Denissenkov}, {Herwig}, {Perdikakis}  \&
  {Schatz}}{{Denissenkov} et~al.}{2021}]{2021MNRAS.503.3913D}
{Denissenkov} P.~A.,  {Herwig} F.,  {Perdikakis} G.,   {Schatz} H.,  2021,
  \mn@doi [\mnras] {10.1093/mnras/stab772}, \href
  {https://ui.adsabs.harvard.edu/abs/2021MNRAS.503.3913D} {503, 3913}

\bibitem[\protect\citeauthoryear{{Diehl} et~al.,}{{Diehl}
  et~al.}{2006}]{2006Natur.439...45D}
{Diehl} R.,  et~al., 2006, \mn@doi [\nat] {10.1038/nature04364}, \href
  {https://ui.adsabs.harvard.edu/abs/2006Natur.439...45D} {439, 45}

\bibitem[\protect\citeauthoryear{{Ebinger}, {Curtis}, {Fr{\"o}hlich}, {Hempel},
  {Perego}, {Liebend{\"o}rfer}  \& {Thielemann}}{{Ebinger}
  et~al.}{2019}]{2019ApJ...870....1E}
{Ebinger} K.,  {Curtis} S.,  {Fr{\"o}hlich} C.,  {Hempel} M.,  {Perego} A.,
  {Liebend{\"o}rfer} M.,   {Thielemann} F.-K.,  2019, \mn@doi [\apj]
  {10.3847/1538-4357/aae7c9}, \href
  {https://ui.adsabs.harvard.edu/abs/2019ApJ...870....1E} {870, 1}

\bibitem[\protect\citeauthoryear{{Ebinger}, {Curtis}, {Ghosh}, {Fr{\"o}hlich},
  {Hempel}, {Perego}, {Liebend{\"o}rfer}  \& {Thielemann}}{{Ebinger}
  et~al.}{2020}]{2020ApJ...888...91E}
{Ebinger} K.,  {Curtis} S.,  {Ghosh} S.,  {Fr{\"o}hlich} C.,  {Hempel} M.,
  {Perego} A.,  {Liebend{\"o}rfer} M.,   {Thielemann} F.-K.,  2020, \mn@doi
  [\apj] {10.3847/1538-4357/ab5dcb}, \href
  {https://ui.adsabs.harvard.edu/abs/2020ApJ...888...91E} {888, 91}

\bibitem[\protect\citeauthoryear{{Ertl}, {Ugliano}, {Janka}, {Marek}  \&
  {Arcones}}{{Ertl} et~al.}{2016}]{2016ApJ...821...69E}
{Ertl} T.,  {Ugliano} M.,  {Janka} H.-T.,  {Marek} A.,   {Arcones} A.,  2016,
  \mn@doi [\apj] {10.3847/0004-637X/821/1/69}, \href
  {https://ui.adsabs.harvard.edu/abs/2016ApJ...821...69E} {821, 69}

\bibitem[\protect\citeauthoryear{{Falla}, {Roberti}, {Limongi}  \&
  {Chieffi}}{{Falla} et~al.}{2025}]{2025ApJ...991...21F}
{Falla} A.,  {Roberti} L.,  {Limongi} M.,   {Chieffi} A.,  2025, \mn@doi [\apj]
  {10.3847/1538-4357/adfa05}, \href
  {https://ui.adsabs.harvard.edu/abs/2025ApJ...991...21F} {991, 21}

\bibitem[\protect\citeauthoryear{{Fields}, {Timmes}, {Farmer}, {Petermann},
  {Wolf}  \& {Couch}}{{Fields} et~al.}{2018}]{2018ApJS..234...19F}
{Fields} C.~E.,  {Timmes} F.~X.,  {Farmer} R.,  {Petermann} I.,  {Wolf} W.~M.,
   {Couch} S.~M.,  2018, \mn@doi [\apjs] {10.3847/1538-4365/aaa29b}, \href
  {https://ui.adsabs.harvard.edu/abs/2018ApJS..234...19F} {234, 19}

\bibitem[\protect\citeauthoryear{{Fr{\"o}hlich} et~al.,}{{Fr{\"o}hlich}
  et~al.}{2006}]{2006ApJ...637..415F}
{Fr{\"o}hlich} C.,  et~al., 2006, \mn@doi [\apj] {10.1086/498224}, \href
  {https://ui.adsabs.harvard.edu/abs/2006ApJ...637..415F} {637, 415}

\bibitem[\protect\citeauthoryear{{Ghosh}, {Wolfe}  \& {Fr{\"o}hlich}}{{Ghosh}
  et~al.}{2022}]{2022ApJ...929...43G}
{Ghosh} S.,  {Wolfe} N.,   {Fr{\"o}hlich} C.,  2022, \mn@doi [\apj]
  {10.3847/1538-4357/ac4d20}, \href
  {https://ui.adsabs.harvard.edu/abs/2022ApJ...929...43G} {929, 43}

\bibitem[\protect\citeauthoryear{{Grefenstette} et~al.,}{{Grefenstette}
  et~al.}{2014}]{2014Natur.506..339G}
{Grefenstette} B.~W.,  et~al., 2014, \mn@doi [\nat] {10.1038/nature12997},
  \href {https://ui.adsabs.harvard.edu/abs/2014Natur.506..339G} {506, 339}

\bibitem[\protect\citeauthoryear{{Heger}, {Fryer}, {Woosley}, {Langer}  \&
  {Hartmann}}{{Heger} et~al.}{2003}]{2003ApJ...591..288H}
{Heger} A.,  {Fryer} C.~L.,  {Woosley} S.~E.,  {Langer} N.,   {Hartmann} D.~H.,
   2003, \mn@doi [\apj] {10.1086/375341}, \href
  {https://ui.adsabs.harvard.edu/abs/2003ApJ...591..288H} {591, 288}

\bibitem[\protect\citeauthoryear{{Hemmerich}}{{Hemmerich}}{2017}]{2017hemmerich}
{Hemmerich} W.~A.,  2017, Korrelation, Korrelationskoeffizient, \url
  {https://web.archive.org/web/20170624012731/http://matheguru.com/stochastik/korrelation.html}

\bibitem[\protect\citeauthoryear{{Hermansen}, {Couch}, {Roberts}, {Schatz}  \&
  {Warren}}{{Hermansen} et~al.}{2020}]{2020ApJ...901...77H}
{Hermansen} K.,  {Couch} S.~M.,  {Roberts} L.~F.,  {Schatz} H.,   {Warren}
  M.~L.,  2020, \mn@doi [\apj] {10.3847/1538-4357/abafb5}, \href
  {https://ui.adsabs.harvard.edu/abs/2020ApJ...901...77H} {901, 77}

\bibitem[\protect\citeauthoryear{{Hirschi} et~al.,}{{Hirschi}
  et~al.}{2025}]{2025MNRAS.543.2796H}
{Hirschi} R.,  et~al., 2025, \mn@doi [\mnras] {10.1093/mnras/staf1470}, \href
  {https://ui.adsabs.harvard.edu/abs/2025MNRAS.543.2796H} {543, 2796}

\bibitem[\protect\citeauthoryear{{Hoffman}, {Woosley}, {Weaver}, {Rauscher}  \&
  {Thielemann}}{{Hoffman} et~al.}{1999}]{1999ApJ...521..735H}
{Hoffman} R.~D.,  {Woosley} S.~E.,  {Weaver} T.~A.,  {Rauscher} T.,
  {Thielemann} F.~K.,  1999, \mn@doi [\apj] {10.1086/307568}, \href
  {https://ui.adsabs.harvard.edu/abs/1999ApJ...521..735H} {521, 735}

\bibitem[\protect\citeauthoryear{{Hoffman} et~al.,}{{Hoffman}
  et~al.}{2010}]{2010ApJ...715.1383H}
{Hoffman} R.~D.,  et~al., 2010, \mn@doi [\apj] {10.1088/0004-637X/715/2/1383},
  \href {https://ui.adsabs.harvard.edu/abs/2010ApJ...715.1383H} {715, 1383}

\bibitem[\protect\citeauthoryear{{Iliadis}, {Champagne}, {Jos{\'e}},
  {Starrfield}  \& {Tupper}}{{Iliadis} et~al.}{2002}]{2002ApJS..142..105I}
{Iliadis} C.,  {Champagne} A.,  {Jos{\'e}} J.,  {Starrfield} S.,   {Tupper} P.,
   2002, \mn@doi [\apjs] {10.1086/341400}, \href
  {https://ui.adsabs.harvard.edu/abs/2002ApJS..142..105I} {142, 105}

\bibitem[\protect\citeauthoryear{{Imasheva}, {Janka}  \& {Weiss}}{{Imasheva}
  et~al.}{2023}]{2023MNRAS.518.1818I}
{Imasheva} L.,  {Janka} H.-T.,   {Weiss} A.,  2023, \mn@doi [\mnras]
  {10.1093/mnras/stac3239}, \href
  {https://ui.adsabs.harvard.edu/abs/2023MNRAS.518.1818I} {518, 1818}

\bibitem[\protect\citeauthoryear{{Imasheva}, {Janka}  \& {Weiss}}{{Imasheva}
  et~al.}{2025}]{2025arXiv250113172I}
{Imasheva} L.,  {Janka} H.~T.,   {Weiss} A.,  2025, \mn@doi [arXiv e-prints]
  {10.48550/arXiv.2501.13172}, \href
  {https://ui.adsabs.harvard.edu/abs/2025arXiv250113172I} {p. arXiv:2501.13172}

\bibitem[\protect\citeauthoryear{{Iyudin} et~al.,}{{Iyudin}
  et~al.}{1994}]{1994A&A...284L...1I}
{Iyudin} A.~F.,  et~al., 1994, \aap, \href
  {https://ui.adsabs.harvard.edu/abs/1994A&A...284L...1I} {284, L1}

\bibitem[\protect\citeauthoryear{Janka}{Janka}{2025}]{2025arXiv250214836J}
Janka H.-T.,  2025, \mn@doi [Annual Review of Nuclear and Particle Science]
  {https://doi.org/10.1146/annurev-nucl-121423-100945}, 75, 425

\bibitem[\protect\citeauthoryear{{Janka}, {Melson}  \& {Summa}}{{Janka}
  et~al.}{2016}]{2016ARNPS..66..341J}
{Janka} H.-T.,  {Melson} T.,   {Summa} A.,  2016, \mn@doi [Annual Review of
  Nuclear and Particle Science] {10.1146/annurev-nucl-102115-044747}, \href
  {https://ui.adsabs.harvard.edu/abs/2016ARNPS..66..341J} {66, 341}

\bibitem[\protect\citeauthoryear{{Kendall}}{{Kendall}}{1955}]{1955kendall}
{Kendall} M.~G.,  1955, {Rank Correlation Methods}.
Charles Griffin, London

\bibitem[\protect\citeauthoryear{{Kobayashi}, {Umeda}, {Nomoto}, {Tominaga}  \&
  {Ohkubo}}{{Kobayashi} et~al.}{2006}]{2006ApJ...653.1145K}
{Kobayashi} C.,  {Umeda} H.,  {Nomoto} K.,  {Tominaga} N.,   {Ohkubo} T.,
  2006, \mn@doi [\apj] {10.1086/508914}, \href
  {https://ui.adsabs.harvard.edu/abs/2006ApJ...653.1145K} {653, 1145}

\bibitem[\protect\citeauthoryear{{Kobayashi}, {Karakas}  \&
  {Lugaro}}{{Kobayashi} et~al.}{2020}]{2020ApJ...900..179K}
{Kobayashi} C.,  {Karakas} A.~I.,   {Lugaro} M.,  2020, \mn@doi [\apj]
  {10.3847/1538-4357/abae65}, \href
  {https://ui.adsabs.harvard.edu/abs/2020ApJ...900..179K} {900, 179}

\bibitem[\protect\citeauthoryear{{Magkotsios}, {Timmes}, {Hungerford}, {Fryer},
  {Young}  \& {Wiescher}}{{Magkotsios} et~al.}{2010}]{2010ApJS..191...66M}
{Magkotsios} G.,  {Timmes} F.~X.,  {Hungerford} A.~L.,  {Fryer} C.~L.,  {Young}
  P.~A.,   {Wiescher} M.,  2010, \mn@doi [\apjs] {10.1088/0067-0049/191/1/66},
  \href {https://ui.adsabs.harvard.edu/abs/2010ApJS..191...66M} {191, 66}

\bibitem[\protect\citeauthoryear{{Martinet}, {Choplin}, {Goriely}  \&
  {Siess}}{{Martinet} et~al.}{2024}]{2024A&A...684A...8M}
{Martinet} S.,  {Choplin} A.,  {Goriely} S.,   {Siess} L.,  2024, \mn@doi
  [\aap] {10.1051/0004-6361/202347734}, \href
  {https://ui.adsabs.harvard.edu/abs/2024A&A...684A...8M} {684, A8}

\bibitem[\protect\citeauthoryear{{M{\"o}ller}, {Nix}, {Myers}  \&
  {Swiatecki}}{{M{\"o}ller} et~al.}{1995}]{1995ADNDT..59..185M}
{M{\"o}ller} P.,  {Nix} J.~R.,  {Myers} W.~D.,   {Swiatecki} W.~J.,  1995,
  \mn@doi [Atomic Data and Nuclear Data Tables] {10.1006/adnd.1995.1002}, \href
  {https://ui.adsabs.harvard.edu/abs/1995ADNDT..59..185M} {59, 185}

\bibitem[\protect\citeauthoryear{{Nagataki}, {Hashimoto}, {Sato}  \&
  {Yamada}}{{Nagataki} et~al.}{1997}]{1997ApJ...486.1026N}
{Nagataki} S.,  {Hashimoto} M.-a.,  {Sato} K.,   {Yamada} S.,  1997, \mn@doi
  [\apj] {10.1086/304565}, \href
  {https://ui.adsabs.harvard.edu/abs/1997ApJ...486.1026N} {486, 1026}

\bibitem[\protect\citeauthoryear{{Nakamura}, {Takiwaki}, {Kotake}  \&
  {Nishimura}}{{Nakamura} et~al.}{2014}]{2014ApJ...782...91N}
{Nakamura} K.,  {Takiwaki} T.,  {Kotake} K.,   {Nishimura} N.,  2014, \mn@doi
  [\apj] {10.1088/0004-637X/782/2/91}, \href
  {https://ui.adsabs.harvard.edu/abs/2014ApJ...782...91N} {782, 91}

\bibitem[\protect\citeauthoryear{{Nishimura}, {Hirschi}, {Rauscher}, {St. J.
  Murphy}  \& {Cescutti}}{{Nishimura} et~al.}{2017}]{2017MNRAS.469.1752N}
{Nishimura} N.,  {Hirschi} R.,  {Rauscher} T.,  {St. J. Murphy} A.,
  {Cescutti} G.,  2017, \mn@doi [\mnras] {10.1093/mnras/stx696}, \href
  {https://ui.adsabs.harvard.edu/abs/2017MNRAS.469.1752N} {469, 1752}

\bibitem[\protect\citeauthoryear{{Nishimura}, {Rauscher}, {Hirschi}, {Murphy},
  {Cescutti}  \& {Travaglio}}{{Nishimura} et~al.}{2018}]{2018MNRAS.474.3133N}
{Nishimura} N.,  {Rauscher} T.,  {Hirschi} R.,  {Murphy} A. S.~J.,  {Cescutti}
  G.,   {Travaglio} C.,  2018, \mn@doi [\mnras] {10.1093/mnras/stx3033}, \href
  {https://ui.adsabs.harvard.edu/abs/2018MNRAS.474.3133N} {474, 3133}

\bibitem[\protect\citeauthoryear{{Nishimura}, {Rauscher}, {Hirschi},
  {Cescutti}, {Murphy}  \& {Fr{\"o}hlich}}{{Nishimura}
  et~al.}{2019}]{2019MNRAS.489.1379N}
{Nishimura} N.,  {Rauscher} T.,  {Hirschi} R.,  {Cescutti} G.,  {Murphy} A.
  S.~J.,   {Fr{\"o}hlich} C.,  2019, \mn@doi [\mnras] {10.1093/mnras/stz2104},
  \href {https://ui.adsabs.harvard.edu/abs/2019MNRAS.489.1379N} {489, 1379}

\bibitem[\protect\citeauthoryear{{Pearson}}{{Pearson}}{1895}]{1895RSPS...58..240P}
{Pearson} K.,  1895, Proceedings of the Royal Society of London Series I, \href
  {https://ui.adsabs.harvard.edu/abs/1895RSPS...58..240P} {58, 240}

\bibitem[\protect\citeauthoryear{{Perego}, {Hempel}, {Fr{\"o}hlich}, {Ebinger},
  {Eichler}, {Casanova}, {Liebend{\"o}rfer}  \& {Thielemann}}{{Perego}
  et~al.}{2015}]{2015ApJ...806..275P}
{Perego} A.,  {Hempel} M.,  {Fr{\"o}hlich} C.,  {Ebinger} K.,  {Eichler} M.,
  {Casanova} J.,  {Liebend{\"o}rfer} M.,   {Thielemann} F.~K.,  2015, \mn@doi
  [\apj] {10.1088/0004-637X/806/2/275}, \href
  {https://ui.adsabs.harvard.edu/abs/2015ApJ...806..275P} {806, 275}

\bibitem[\protect\citeauthoryear{{Pruet}, {Hoffman}, {Woosley}, {Janka}  \&
  {Buras}}{{Pruet} et~al.}{2006}]{2006ApJ...644.1028P}
{Pruet} J.,  {Hoffman} R.~D.,  {Woosley} S.~E.,  {Janka} H.~T.,   {Buras} R.,
  2006, \mn@doi [\apj] {10.1086/503891}, \href
  {https://ui.adsabs.harvard.edu/abs/2006ApJ...644.1028P} {644, 1028}

\bibitem[\protect\citeauthoryear{{Rauscher}}{{Rauscher}}{2012}]{2012ApJS..201...26R}
{Rauscher} T.,  2012, \mn@doi [\apjs] {10.1088/0067-0049/201/2/26}, \href
  {https://ui.adsabs.harvard.edu/abs/2012ApJS..201...26R} {201, 26}

\bibitem[\protect\citeauthoryear{{Rauscher}}{{Rauscher}}{2020a}]{2020entn.book.....R}
{Rauscher} T.,  2020a, {Essentials of Nucleosynthesis and Theoretical Nuclear
  Astrophysics}, \mn@doi{10.1088/2514-3433/ab8737.
}

\bibitem[\protect\citeauthoryear{{Rauscher}}{{Rauscher}}{2020b}]{2020JPhCS1643a2062R}
{Rauscher} T.,  2020b, in Journal of Physics Conference Series. IOP, p. 012062
  (\mn@eprint {arXiv} {1909.06830}), \mn@doi{10.1088/1742-6596/1643/1/012062}

\bibitem[\protect\citeauthoryear{{Rauscher} \& {Thielemann}}{{Rauscher} \&
  {Thielemann}}{2000}]{2000ADNDT..75....1R}
{Rauscher} T.,  {Thielemann} F.-K.,  2000, \mn@doi [Atomic Data and Nuclear
  Data Tables] {10.1006/adnd.2000.0834}, \href
  {https://ui.adsabs.harvard.edu/abs/2000ADNDT..75....1R} {75, 1}

\bibitem[\protect\citeauthoryear{{Rauscher}, {Heger}, {Hoffman}  \&
  {Woosley}}{{Rauscher} et~al.}{2002}]{2002ApJ...576..323R}
{Rauscher} T.,  {Heger} A.,  {Hoffman} R.~D.,   {Woosley} S.~E.,  2002, \mn@doi
  [\apj] {10.1086/341728}, \href
  {https://ui.adsabs.harvard.edu/abs/2002ApJ...576..323R} {576, 323}

\bibitem[\protect\citeauthoryear{{Rauscher}, {Mohr}, {Dillmann}  \&
  {Plag}}{{Rauscher} et~al.}{2011}]{2011ApJ...738..143R}
{Rauscher} T.,  {Mohr} P.,  {Dillmann} I.,   {Plag} R.,  2011, \mn@doi [\apj]
  {10.1088/0004-637X/738/2/143}, \href
  {https://ui.adsabs.harvard.edu/abs/2011ApJ...738..143R} {738, 143}

\bibitem[\protect\citeauthoryear{{Rauscher}, {Nishimura}, {Hirschi},
  {Cescutti}, {Murphy}  \& {Heger}}{{Rauscher}
  et~al.}{2016}]{2016MNRAS.463.4153R}
{Rauscher} T.,  {Nishimura} N.,  {Hirschi} R.,  {Cescutti} G.,  {Murphy} A.
  S.~J.,   {Heger} A.,  2016, \mn@doi [\mnras] {10.1093/mnras/stw2266}, \href
  {https://ui.adsabs.harvard.edu/abs/2016MNRAS.463.4153R} {463, 4153}

\bibitem[\protect\citeauthoryear{{Rauscher}, {Nishimura}, {Cescutti},
  {Hirschi}, {Murphy}  \& {Travaglio}}{{Rauscher}
  et~al.}{2020}]{2020nnc..confa0061R}
{Rauscher} T.,  {Nishimura} N.,  {Cescutti} G.,  {Hirschi} R.,  {Murphy} A.
  S.~J.,   {Travaglio} C.,  2020, in JPS Conference Proceedings. The Physical
  Society of Japan, p. 010061 (\mn@eprint {arXiv} {1907.09178}),
  \mn@doi{10.7566/JPSCP.32.010061}

\bibitem[\protect\citeauthoryear{{Reichert} et~al.,}{{Reichert}
  et~al.}{2023}]{2023ApJS..268...66R}
{Reichert} M.,  et~al., 2023, \mn@doi [\apjs] {10.3847/1538-4365/acf033}, \href
  {https://ui.adsabs.harvard.edu/abs/2023ApJS..268...66R} {268, 66}

\bibitem[\protect\citeauthoryear{{Sawada} \& {Suwa}}{{Sawada} \&
  {Suwa}}{2023}]{2023arXiv230103610S}
{Sawada} R.,  {Suwa} Y.,  2023, \mn@doi [arXiv e-prints]
  {10.48550/arXiv.2301.03610}, \href
  {https://ui.adsabs.harvard.edu/abs/2023arXiv230103610S} {p. arXiv:2301.03610}

\bibitem[\protect\citeauthoryear{{Siegert}, {Diehl}, {Krause}  \&
  {Greiner}}{{Siegert} et~al.}{2015}]{2015A&A...579A.124S}
{Siegert} T.,  {Diehl} R.,  {Krause} M. G.~H.,   {Greiner} J.,  2015, \mn@doi
  [\aap] {10.1051/0004-6361/201525877}, \href
  {https://ui.adsabs.harvard.edu/abs/2015A&A...579A.124S} {579, A124}

\bibitem[\protect\citeauthoryear{{Sonzogni} et~al.,}{{Sonzogni}
  et~al.}{2000}]{2000PhRvL..84.1651S}
{Sonzogni} A.~A.,  et~al., 2000, \mn@doi [\prl] {10.1103/PhysRevLett.84.1651},
  \href {https://ui.adsabs.harvard.edu/abs/2000PhRvL..84.1651S} {84, 1651}

\bibitem[\protect\citeauthoryear{{Steiner}, {Hempel}  \& {Fischer}}{{Steiner}
  et~al.}{2013}]{2013ApJ...774...17S}
{Steiner} A.~W.,  {Hempel} M.,   {Fischer} T.,  2013, \mn@doi [\apj]
  {10.1088/0004-637X/774/1/17}, \href
  {https://ui.adsabs.harvard.edu/abs/2013ApJ...774...17S} {774, 17}

\bibitem[\protect\citeauthoryear{{Subedi}, {Meisel}  \& {Merz}}{{Subedi}
  et~al.}{2020}]{2020ApJ...898....5S}
{Subedi} S.~K.,  {Meisel} Z.,   {Merz} G.,  2020, \mn@doi [\apj]
  {10.3847/1538-4357/ab9745}, \href
  {https://ui.adsabs.harvard.edu/abs/2020ApJ...898....5S} {898, 5}

\bibitem[\protect\citeauthoryear{{The}, {Clayton}, {Jin}  \& {Meyer}}{{The}
  et~al.}{1998}]{1998ApJ...504..500T}
{The} L.~S.,  {Clayton} D.~D.,  {Jin} L.,   {Meyer} B.~S.,  1998, \mn@doi
  [\apj] {10.1086/306057}, \href
  {https://ui.adsabs.harvard.edu/abs/1998ApJ...504..500T} {504, 500}

\bibitem[\protect\citeauthoryear{{Thielemann}, {Nomoto}  \&
  {Hashimoto}}{{Thielemann} et~al.}{1996}]{1996ApJ...460..408T}
{Thielemann} F.-K.,  {Nomoto} K.,   {Hashimoto} M.-A.,  1996, \mn@doi [\apj]
  {10.1086/176980}, \href
  {https://ui.adsabs.harvard.edu/abs/1996ApJ...460..408T} {460, 408}

\bibitem[\protect\citeauthoryear{{Timmes} et~al.,}{{Timmes}
  et~al.}{2019}]{2019BAAS...51c...2T}
{Timmes} F.,  et~al., 2019, \mn@doi [\baas] {10.48550/arXiv.1902.02915}, \href
  {https://ui.adsabs.harvard.edu/abs/2019BAAS...51c...2T} {51, 2}

\bibitem[\protect\citeauthoryear{{Tur}, {Heger}  \& {Austin}}{{Tur}
  et~al.}{2007}]{2007ApJ...671..821T}
{Tur} C.,  {Heger} A.,   {Austin} S.~M.,  2007, \mn@doi [\apj]
  {10.1086/523095}, \href
  {https://ui.adsabs.harvard.edu/abs/2007ApJ...671..821T} {671, 821}

\bibitem[\protect\citeauthoryear{{Ugliano}, {Janka}, {Marek}  \&
  {Arcones}}{{Ugliano} et~al.}{2012}]{2012ApJ...757...69U}
{Ugliano} M.,  {Janka} H.-T.,  {Marek} A.,   {Arcones} A.,  2012, \mn@doi
  [\apj] {10.1088/0004-637X/757/1/69}, \href
  {https://ui.adsabs.harvard.edu/abs/2012ApJ...757...69U} {757, 69}

\bibitem[\protect\citeauthoryear{{Wanajo}}{{Wanajo}}{2006}]{2006ApJ...647.1323W}
{Wanajo} S.,  2006, \mn@doi [\apj] {10.1086/505483}, \href
  {https://ui.adsabs.harvard.edu/abs/2006ApJ...647.1323W} {647, 1323}

\bibitem[\protect\citeauthoryear{{Wang} \& {Burrows}}{{Wang} \&
  {Burrows}}{2024}]{2024ApJ...974...39W}
{Wang} T.,  {Burrows} A.,  2024, \mn@doi [\apj] {10.3847/1538-4357/ad6983},
  \href {https://ui.adsabs.harvard.edu/abs/2024ApJ...974...39W} {974, 39}

\bibitem[\protect\citeauthoryear{{Wongwathanarat}, {Janka}, {M{\"u}ller},
  {Pllumbi}  \& {Wanajo}}{{Wongwathanarat} et~al.}{2017}]{2017ApJ...842...13W}
{Wongwathanarat} A.,  {Janka} H.-T.,  {M{\"u}ller} E.,  {Pllumbi} E.,
  {Wanajo} S.,  2017, \mn@doi [\apj] {10.3847/1538-4357/aa72de}, \href
  {https://ui.adsabs.harvard.edu/abs/2017ApJ...842...13W} {842, 13}

\bibitem[\protect\citeauthoryear{{Woosley} \& {Weaver}}{{Woosley} \&
  {Weaver}}{1995}]{1995ApJS..101..181W}
{Woosley} S.~E.,  {Weaver} T.~A.,  1995, \mn@doi [\apjs] {10.1086/192237},
  \href {https://ui.adsabs.harvard.edu/abs/1995ApJS..101..181W} {101, 181}

\bibitem[\protect\citeauthoryear{{Yamada} et~al.,}{{Yamada}
  et~al.}{2024}]{2024PJAB..100..190Y}
{Yamada} S.,  et~al., 2024, \mn@doi [Proceedings of the Japan Academy, Series
  B] {10.2183/pjab.100.015}, \href
  {https://ui.adsabs.harvard.edu/abs/2024PJAB..100..190Y} {100, 190}

\makeatother
\end{thebibliography}





\bsp	
\label{lastpage}
\end{document}